\documentclass{ieeeaccess}
\usepackage{cite}
\usepackage{amsmath,amssymb,amsfonts}
\usepackage{algorithmic}
\usepackage{graphicx}
\usepackage{textcomp}
\usepackage{etoolbox}
\usepackage{graphics}
\usepackage{fontawesome}
\usepackage{multirow}
\usepackage{longtable}
\usepackage{pdfpages}
\usepackage{afterpage}
\usepackage{pdflscape}
\usepackage{array}

\usepackage{cleveref}
\usepackage{tabularx}
\usepackage{booktabs}
\usepackage{multirow}
\newcolumntype{C}[1]{>{\centering\arraybackslash\hspace{0pt}}p{#1}}
\newcolumntype{L}[1]{>{\raggedright\arraybackslash\hspace{0pt}}p{#1}}
\usepackage[skip=1ex]{caption}
\usepackage{booktabs, tabularx}
\usepackage{enumitem}
\usepackage{etoolbox}
\AtBeginEnvironment{table}{%
\setlist[itemize]{nosep,
                 leftmargin=*,
                 label=\textbullet,
                 before=\begin{minipage}[t]{\linewidth}, 
                 after=\end{minipage}}                   
                        }
\newcommand{\todo}[1]{}
\renewcommand{\todo}[1]{{\color{red} TODO: {#1}}}

\AtBeginDocument{%
  \providecommand\BibTeX{{%
    \normalfont B\kern-0.5em{\scshape i\kern-0.25em b}\kern-0.8em\TeX}}}

\def\BibTeX{{\rm B\kern-.05em{\sc i\kern-.025em b}\kern-.08em
    T\kern-.1667em\lower.7ex\hbox{E}\kern-.125emX}}
\begin{document}
\history{Date of publication xxxx 00, 0000, date of current version xxxx 00, 0000.}
\doi{AAAAA}

\title{Operationalizing Human Values in Software Engineering: A Survey}
\author{\uppercase{Mojtaba Shahin}\authorrefmark{1}, \IEEEmembership{Member, IEEE},
\uppercase{Waqar Hussain\authorrefmark{2}, \IEEEmembership{Member, IEEE}, Arif Nurwidyantoro\authorrefmark{3}, \IEEEmembership{Member, IEEE}, Harsha Perera\authorrefmark{4},\IEEEmembership{Member, IEEE}, Rifat Shams \authorrefmark{5},\IEEEmembership{Member, IEEE}, John Grundy \authorrefmark{6},\IEEEmembership{Member, IEEE}, and Jon Whittle \authorrefmark{7}},
\IEEEmembership{Member, IEEE}}
\address[1]{RMIT University, Melbourne, Australia (e-mail: mojtaba.shahin@rmit.edu.au)}
\address[2]{CSIRO's Data61, Australia (e-mail: waqar.hussain@data61.csiro.au)}
\address[3]{Monash Univeristy, Melbourne, Australia (e-mail: arif.nurwidyantoro@monash.edu)}
\address[4]{Monash Univeristy, Melbourne, Australia (e-mail: harsha.perera@monash.edu)}
\address[5]{Monash Univeristy, Melbourne, Australia (e-mail: rifat.shams@monash.edu)}
\address[6]{Monash Univeristy, Melbourne, Australia (e-mail: john.grundy@monash.edu)}
\address[7]{CSIRO's Data61, Australia (e-mail: jon.whittle@data61.csiro.au)}

\markboth
{Shahin \headeretal: Operationalizing Human Values in Software Engineering: A Survey}
{Shahin \headeretal: Operationalizing Human Values in Software Engineering: A Survey}

\corresp{Corresponding author: Mojtaba Shahin (e-mail: mojtaba.shahin@rmit.edu.au).}

\begin{abstract}
Human values (e.g., pleasure, privacy, and social justice) are what a person or a society considers important. The inability to address them in software-intensive systems can result in numerous undesired consequences (e.g., financial losses) for individuals and communities. Various solutions (e.g., methodologies, techniques) are developed to help ``operationalize values in software''. The ultimate goal is to ensure building software (better) reflects and respects human values. In this survey, ``operationalizing values'' is referred to as \textit{the process of identifying human values and translating them to accessible and concrete concepts so that they can be implemented, validated, verified, and measured in software}. This paper provides a deep understanding of the research landscape on operationalizing values in software engineering, covering 51 primary studies. It also presents an analysis and taxonomy of 51 solutions for operationalizing values in software engineering. Our survey reveals that most solutions attempt to help operationalize values in the early phases (requirements and design) of the software development life cycle. However, the later phases (implementation and testing) and other aspects of software development (e.g., ``team organization'') still need adequate consideration. We outline implications for research and practice and identify open issues and future research directions to advance this area.

\end{abstract}


\titlepgskip=-15pt

\maketitle
\section{Introduction}
Software systems (e.g., mobile apps, banking systems, video games) are now an integrated part of our society. Software systems are expected to address, respect, and be aligned with the individual human values of their diverse end-users who might have different characteristics (e.g., aging people, visually challenged people) \cite{whittle2021case, ferrario2016values, sellen2009reflecting}. On the macro level, software systems should not harm or jeopardize social justice and human rights (such as privacy) \cite{kirkham2020using}. Human values such as inclusion, diversity, autonomy, and wealth are defined as ``what is important for an individual or a society'' \cite{schwartz2012overview}. Human values are also referred to as the principles that guide human actions and behavior in daily life \cite{rokeach1973nature}. Failing to address human values in software-intensive systems may bring problems and irreversible damages for all stakeholders who are directly or indirectly influenced by software-intensive systems \cite{koch2013approximate,galhotra2017fairness}. These difficulties and damages are enormous and can range from user dissatisfaction to reputational disaster, financial losses, or loss of life.

Some examples of human values ignored or violated by software systems and their creators have had such devastating and widespread damages that they have been widely covered in the media and led to public condemnation of the software industry \cite{galhotra2017fairness, incidentDB}. For example, Facebook and Cambridge Analytica were accused of violating \textit{privacy} and abusing \textit{power} by harvesting and using almost 87 million Facebook users' personal data without seeking their consent to help influence voters’ choices in the US presidential election \cite{fbCamAna}. Facebook faced large fines (i.e., US\$ 5 billion) and lost US\$ 119 billion stock value in one day \cite{neate_2018}. Amazon's ‘‘Prime same-day delivery service’’, which is designed to provide an egalitarian shopping experience for all US citizens, appears to be \textit{unfair} and \textit{biased} against black neighborhoods as they are systematically deprived of receiving this service \cite{Amazon_2016}.


Operationalizing human values in software is expected to prevent or minimize such undesired effects and issues  and bring benefits for software organizations, end-users, and practitioners such as an excellent reputation for software organizations, end-users put trust in the software, and the increased acceptability of the software \cite{wang2013affects, whittle2021case}. Inspired by the literature \cite{nissenbaum2011preemption, flanagan2007game, mougouei2018operationalizing}, we define \textbf{``operationalizing human values in software''} as \textit{the process of identifying human values and translating them to accessible and concrete concepts so that they can be implemented, validated, verified, and measured in software}.          

Due to the increasing importance of operationalizing human values in software, a growing body of literature has attempted to provide solutions (e.g., frameworks, tools, roles, design patterns, etc.) to help operationalize human values in software. However, such solutions and the knowledge (e.g., their motivations and limitations) around them are scattered in the literature that appears in diverse venues. Consequently, there is no clear and holistic view on how human values can be operationalized in software. We expect that having a comprehensive understanding of operationalizing values in software helps identify the areas such as tooling support and methodological aspects which need more support and investments. Further to this, such a comprehensive understanding can help software development organizations become more aware of possible solutions and associated tools for operationalizing values and adopt appropriate ones that match their needs and industrial settings.

A few secondary studies have reviewed the literature on human values \cite{Friedman2017,Lenberg2015} and human values in software \cite{Perera2020, Salleh2019,Khurum2013}. Perera et al. \cite{Perera2020} investigated to what extent papers in the leading software engineering venues are relevant to values. Friedman et al. \cite{Friedman2017} studied 14 methods that aim to consider and integrate values in the design process of technologies. Salleh et al. \cite{Salleh2019} and Khurum et al. \cite{Khurum2013} carried out systematic mapping studies on Value-based Software Engineering (VBSE). VBSE mainly sees the concept of ‘value’ from the economic lens \cite{biffl2006value}. While Salleh et al. \cite{Salleh2019} focused on characterizing the research around VBSE (e.g., principles, research methods, etc.), Khurum et al. \cite{Khurum2013} developed the Software Value Map (SVM) to provide a unified and consolidated view of value. Our survey differs from these review studies in terms of objectives, the level of in-depth analysis, and research questions (See Section \ref{sec:existingreviws}). The scope of our survey is identifying, analyzing, and classifying solutions for operationalizing human values applied to software-intensive systems development. None of the previous works focused on this aspect. Our survey does not focus on operationalizing human values in technology or product development (e.g., car design).


To gain a comprehensive understanding of the state-of-the-art solutions for operationalizing values in software, we conducted a survey on 51 primary studies. We found these 51 primary studies by following Webster and Watson's guidelines \cite{webster2002analyzing}, suggesting identifying a pool of initial papers, followed by applying the backward snowballing technique.
(1) Our survey identifies 51 solutions to operationalize values, which can contribute to five areas in software engineering: \textit{requirements}, \textit{design}, \textit{implementation}, \textit{testing}, and \textit{team organization}. (2) The 51 solutions are further classified into 10 ``not mutually exclusive'' groups, in which the majority of them (31) attempt to ``capture values from different resources (e.g., stakeholders)''. (3) The majority of solutions (32, 62.7\%) are able to operationalize any values, while the rest (19) target one or two exclusive values. (4) Only a few solutions (14 out 51, 27.4\%) are supported by (semi-) automated tools.    

The key contributions of this survey are
\begin{itemize}
    \item The first comprehensive survey on the current research on operationalizing values in software;
    \item A taxonomy of solutions published to 2020 for operationalizing values in software;
    \item A list of promising research directions for future work and investments;
\end{itemize}
 In Section \ref{sec:Background}, we provide some definitions of human values and introduce some well-known values models. Section \ref{sec:existingreviws} summarizes existing review studies on human values. Section \ref{sec:researchmethod} outlines our research method. We report our findings in Sections \ref{sec:demographics}, \ref{sec:taxonomy}, and \ref{sec:solutionsanalysis}. Section \ref{sec:discussion} reflects on the key findings and proposes some promising research areas. In Section \ref{limiations}, we discuss possible limitations and threats to the validity of our survey. Finally, we conclude our paper in Section \ref{sec:conclusion}.
\section{Background}\label{sec:Background}
\subsection{Human Value Definitions}\label{sec:Definitions}
The concept of ``human values'' has long been of interest among the researchers of sociology, psychology, anthropology \cite{schwartz2012overview} as well as science and engineering \cite{Perera2020, shams2020}. As defined by Schwartz, \textit{``human values are desirable, trans-situational goals, varying in importance, that serve as guiding principles in people’s lives''} \cite{schwartz1996value}. Therefore, human values are something that individuals deem important in life \cite{rokeach1973nature}. Values can be fundamental and primary needs (e.g., food) or general needs (e.g., self-esteem) \cite{gouveia2014functional}. Many researchers from social science defined values as abstract goals, individual attitudes, behaviors, and beliefs \cite{parashar2004perception}. Schwartz and Bilsky define values as \textit{``(a) concepts or beliefs, (b) about desirable end states or behaviors, (c) that transcend specific situations, (d) guide selection or evaluation of behavior and events, and (e) are ordered by relative importance''} \cite{schwartz1987toward}. Values are also defined as principles that guide social life and are modes of conduct that a person likes or chooses among different situations \cite{parashar2004perception, rokeach1973nature}. Above all, values are ``ways to live'' \cite{rohan2000rose} which can be defined as a micro-macro concept where the micro level is individual behavior and macro level is cultural practices \cite{parashar2004perception}. 

Related to values are two important concepts from human psychology, namely human motivation and emotions, that are worth mentioning. Motivation serves as a guiding force for all human behavior and actions. Humans are goal-directed creatures, and their motivation energizes, directs, and sustains their goal-directed activities \cite{schunk2008motivation}. The sought-out goals can be as concrete as obtaining food or clothing or abstract like developing a sense of meaning or purpose. On the other hand, emotions are kinds of desires, action tendencies, or feelings that correspond to physiological changes brought on by pleasure/displeasure or behavioral responses, e.g., heart racing when looking at an object perceived as dangerous.

Emotions are often intertwined with personality and motivation; at times aligned with motivational goals and rationality but often challenging any practical reasons \cite{tappolet2016emotions}.
According to modern theories of motivation, values and emotions underlie human motivation. The inter-relationship of these concepts helps us understand why individuals behave in the manner they do, e.g., approve or disapprove of something and engage or disengage in various activities \cite{eccles2002motivational}. While some values do have a moral import, not all values are derived from ethics or moral philosophy; the societal or religious perceptions of what is morally acceptable or unacceptable behavior \cite{fieser2016ethics}.

\subsection{Values Models}\label{sec:valuesmoels}
\begin{figure*}
    \centering
    \includegraphics[width=0.70\textwidth]{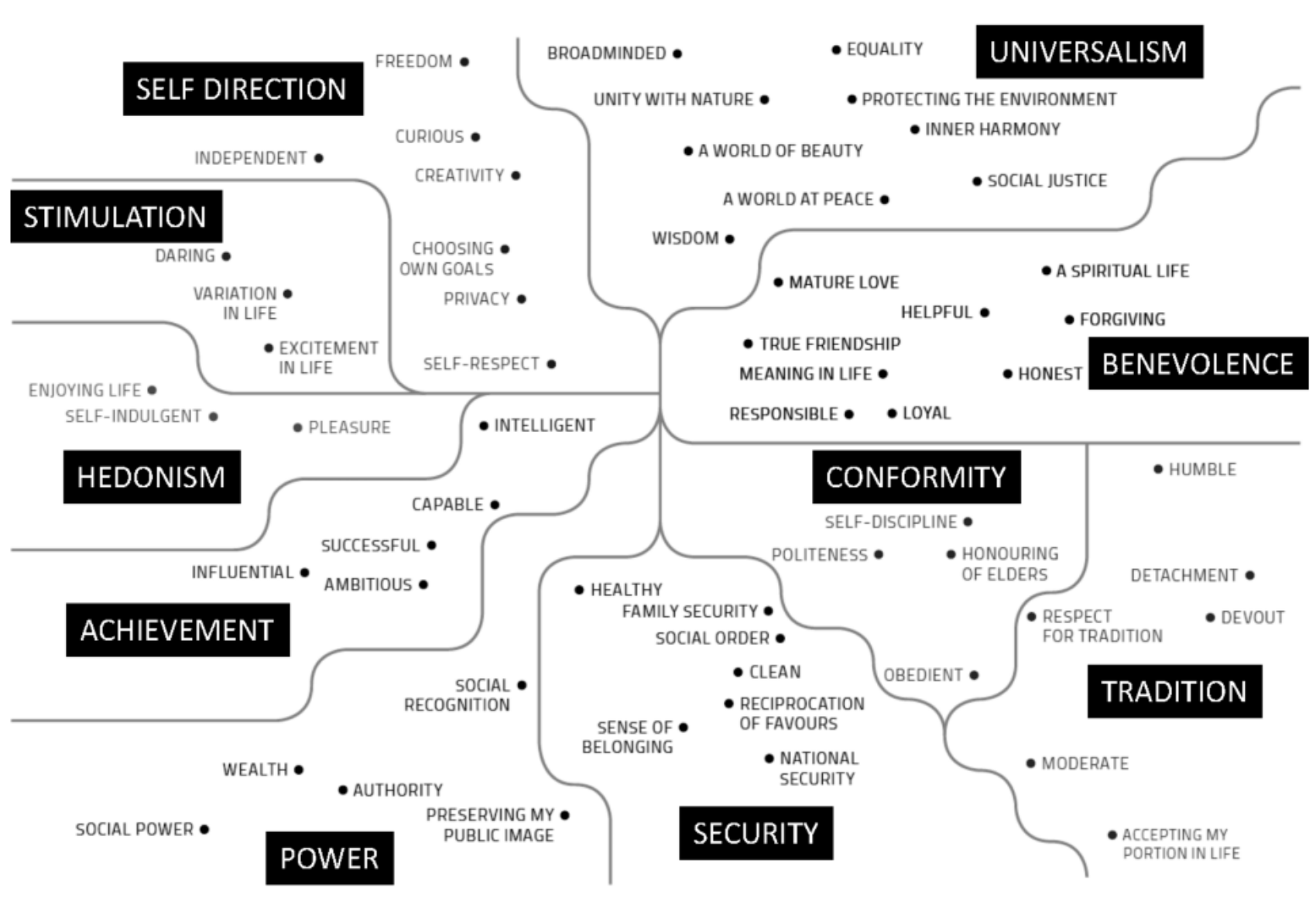}
    \caption{Schwartz values model \cite{schwartz1992universals} (adopted from \cite{CommonCauseHandbook,ferrario2016values}). The boxes indicate the ten universal values, and each of them is subdivided into some finer-grained values. 
    }
    \label{fig:SchwartzModel}
\end{figure*}

There is no universal agreement on either the number of human values or the way human values can be modeled. Nevertheless, several human values models introduced in social sciences are recognized as the most comprehensive values models to date. Table \ref{tbl:vmodels} provides an overview of six well-known human values models. In 1973, Rokeach identified 36 universal human values using a survey-based approach. Half of the values introduced in Rokeach's model are `life goals' such as \textit{Inner Harmony} and \textit{Social Recognition}, while the rest are linked to modes of behavior such as \textit{Cheerfulness} and \textit{Politeness} \cite{rokeach1973nature}. Taking forward the survey-based approach, Schwartz (1992) proposed the theory of basic human values  \cite{schwartz1992universals}. The theory includes ten main value categories, which are measured with 58 individual value items. 
Importantly, as illustrated in Figure \ref{fig:SchwartzModel}, Schwartz organized values in a circular structure to depict their relationships. Values (e.g., \textit{Self-Direction} and \textit{Stimulation}) that appear close to each other are complementary, and those (e.g., \textit{Self-Direction} vs. \textit{Tradition}) that are further apart are in conflict. This theory has been developed using data from 82 countries with different socio-ethnic backgrounds \cite{schwartz1992universals}. 

 \begin{table*}[ht]
    \caption{Six well-known values models.}
    \footnotesize
\label{tbl:vmodels}
\begin{tabularx}{
\linewidth}{l|l|l|l|l|X}

    \toprule
                  
      \textbf{Model} & \textbf{Authors} & \textbf{Year}
            & \textbf{Num. of Values} & \textbf{Method} & \textbf{Notes}          \\ 
    \toprule
    
\begin{tabular}[c]{@{}l@{}} Rokeach’s Value\\ Survey  \cite{rokeach1973nature}   \end{tabular}                                                  & M. Rokeach & 1973 & \begin{tabular}[c]{@{}l@{}} 36 values grouped\\ in 2 categories \end{tabular}                & Survey                                                & The model proposes a theoretical connection between values and behaviors.                                                           \\ \hline
\begin{tabular}[c]{@{}l@{}} Culture's\\ Consequences \cite{hofstede1984hofstede} \end{tabular}                                                     & \begin{tabular}[c]{@{}l@{}} G. Hofstede,\\ M.H. Bond
\end{tabular}  & 1984 &  \begin{tabular}[c]{@{}l@{}}4 cultural\\ dimensions        \end{tabular}             & Survey                                                & The proposed dimensions can be used to distinguish one culture from another.                                                          \\ \hline
List of Values \cite{kahle1988using}                                                             & \begin{tabular}[c]{@{}l@{}} L.R. Kahle,\\ P. Kennedy \end{tabular}  & 1988 & 9 values                                                                                                      & \begin{tabular}[c]{@{}l@{}}Literature\\ Review\end{tabular} & The model is well known in advertising and marketing and focuses on the values that are visible in day-to-day lives. 
\\ \hline
\begin{tabular}[c]{@{}l@{}} Schwartz Theory \\of Basic\\ Human Values \cite{schwartz1992universals} \end{tabular}  & S.H. Schwartz & 1992 & \begin{tabular}[c]{@{}l@{}}58 values grouped\\ in 10 categories \end{tabular} 
& Survey                                                & The model has been empirically tested in 82 countries. Arguably these values are considered universal human values. \\ \hline
\begin{tabular}[c]{@{}l@{}} Meta-inventory of\\ Human Values \cite{cheng2010developing}\end{tabular}   & \begin{tabular}[c]{@{}l@{}} A. Cheng,\\ K.R. Fleischmann \end{tabular}  & 2010 & 16 values                                                                                                   & \begin{tabular}[c]{@{}l@{}}Literature\\ Review\end{tabular} & It is a meta-inventory based on 12 value inventories from the literature.                                                                                                        \\ \hline
\begin{tabular}[c]{@{}l@{}} Functional Theory\\ of Human Values \cite{gouveia2014functional}\end{tabular} & \begin{tabular}[c]{@{}l@{}} V.V. Gouveia,\\T.L. Milfont,\\V.M. Guerra \end{tabular} & 2014 & \begin{tabular}[c]{@{}l@{}} 18 values grouped\\ in 6 categories  \end{tabular}                            & Survey                                                & The two dimensions of the matrix represent values regarding human behavior and human needs.                                          \\ 
    
  \bottomrule

\end{tabularx}%
    \end{table*}
Hofstede used value measurement analysis mainly on cultural aspects and suggested four cultural dimensions: \textit{``Power Distance''}, \textit{``Uncertainty Avoidance''}, \textit{``Individualism versus Collectivism''}, and \textit{``Masculinity versus Femininity''} \cite{hofstede1984hofstede}. 
Moreover, there are noteworthy contributions to human values models proposed by various researchers. 
For example, Parashar et al. (2004) introduced the micro and macro concept of values which are individual behavior and cultural practices, respectively \cite{parashar2004perception}. Gouveia et al. introduced the three-by-two framework with six basic value categories, and three specific values under each category \cite{gouveia2014functional}. 
Cheng and Fleischmann reviewed 12 different values models from different disciplines to produce a meta-inventory of values \cite{cheng2010developing}. They categorized all human values from these 12 models into 16 main values categories. Their categorization allows us to understand the similar values concepts (or synonyms) discussed using different wordings in different values models. 
For example, to describe life's accomplishments, Schwartz uses the word \textit{Achievement} while Rockeach uses \textit{A Sense of Accomplishment}. The same idea is expressed by Kahle et al. \cite{kahle1988using} as \textit{Self-fulfillment}. 

The primary studies investigated in our survey use different values models as a reference point to motivate, describe, and evaluate their proposed solutions. Consequently, they may use various terminology to refer to the same human value. We carefully considered such similarities in our study and selected a common terminology.


\section{Existing Reviews on Human Values}\label{sec:existingreviws}



There are a few review papers on human values in software engineering. Perera et al. investigated the abstract of 1350 papers published between 2015 and 2018 in four top-tier software engineering venues to identify which ones are relevant to human values using the Schwartz theory of basic values ~\cite{Perera2020}. Perera et al. found that only 216 papers (16\%) out of the 1350 papers are directly relevant to human values (e.g., leveraging human values as the main driver in developers' decisions). The study concluded that apart from a few values such as \textit{Security}, \textit{Privacy}, and \textit{Helpful}, other human values in the Schwartz theory, such as \textit{Curiosity}, \textit{Pleasure}, and \textit{Social Justice}, are inadequately addressed to date in software engineering research. 
Unlike their study, in our survey:

\begin{enumerate}
    \item Our focus is to find solutions proposed in the literature for operationalizing human values in software engineering;
    \item We develop a taxonomy of solutions for incorporating human values in software;
    \item We provide comprehensive contextual information (e.g., research types and methodologies) of the studies that attempt to provide solutions to embed human values in software.

\end{enumerate}

Value-based Software Engineering (VBSE) is one of the earliest attempts to address values in software engineering. VBSE was proposed as a paradigm shift from the traditional \textit{`value-neutral'} towards a value-based approach for developing software~\cite{Boehm2003}. In regards to this, Salleh et al. performed a systematic mapping study to understand the state of VBSE research published between 2003 to 2017~\cite{Salleh2019}. 
The study classified 134 papers based on software engineering principles and practices, research methodologies, and research types. The study results show that: (1) the two leading software engineering principles and practices ~\cite{Boehm2003} explored in the VBSE community are \textit{Value-based (VB) requirements engineering} and \textit{VB planning and control}, (2) case study was the main methodology used in VBSE research, and (3) previous VBSE studies mostly proposed solutions without any validation. In contrast to human values, VBSE is chiefly concerned with economic value. Furthermore, dissimilar to their study, we developed our own taxonomy of solutions to incorporate human values and investigated if the proposed solutions address specific human values, such as \textit{fairness} or \textit{authority}.


Khurum et al.~\cite{Khurum2013} carried out a systematic mapping to discover value-based aspects relevant to decision-making in software-intensive product development. Consequently, they introduced \textit{``Software Value Map''} (SVM) as a consolidated view of the concept of value aggregated from the finance, customer, business process, and innovation/learning perspectives. Apart from building a unified view of value, they made two other notable contributions relevant to our discussion of value here: (1) categorizing value constructs as \textit{``value aspects''}, \textit{``value sub-aspects''}, and \textit{``value components''}, to facilitate the development of shared understanding among decision-makers and (2) mapping the interrelationship amongst various value constructs explicitly and collecting various methods to measure a specific value component.  On their part, this was an attempt to facilitate practitioners' understanding of value for one or more perspectives and enhance their ability to make an informed decision about value creation in the (software) product they produce. Although customer perspective and their intrinsic value were explored, almost none of the human values such as \textit{benevolence}, \textit{universalism}, \textit{self-direction} mentioned in the values models introduced in Section \ref{sec:valuesmoels} were addressed. Furthermore, no solutions were identified to explicitly address or measure them.


Value Sensitive Design (VSD) is a method to incorporate values during the design phase of a product. VSD is defined as \textit{``a theoretically grounded approach to the design of technology that accounts for human values in a principled and systematic manner throughout the design process''}~\cite{Friedman2017}.
Friedman et al.~\cite{Friedman2017} identified and reviewed the existing VSD methods proposed in the literature by applying the following inclusion criteria: (1) the method has been invented or undergone substantial development for the investigation of values in technology, (2) the method is self-contained, and (3) the method covers a broad range of values and application areas. The use of these selection criteria resulted in a collection of 14 VSD methods, such as \textit{value source analysis} and \textit{value scenario}.
With respect to their study, we argue that the definition of VSD is limited to only the design process, but on the other hand, also broader by covering the development of any technology. For example, one VSD method called \textit{value sketch} focused on \textit{`understandings, views, and values about a technology'}~\cite{Friedman2017}, not specific about software. Our study aims to complement Friedman et al.'s study by covering solutions proposed in the literature to operationalize values in all software engineering aspects.

Another concept named `Behavioral Software Engineering (BSE)' is, to some extent, relevant to human values~\cite{Lenberg2015}. Lenberg et al. defined BSE as \textit{``the study of cognitive, behavioral, and social aspects of software engineering performed by individuals, groups, or organizations''}~\cite{Lenberg2015}. Lenberg et al. reviewed 250 papers published between 1997 and 2013 that discussed BSE concepts in the software engineering discipline ~\cite{Lenberg2015}. They found that the software engineering research did not (sufficiently) investigate the majority of BSE concepts (e.g., \textit{life satisfaction}, \textit{conformity}) and only focused on a few BSE concepts (e.g., \textit{communication}, \textit{personality}, and \textit{group composition}). It was also found that the vast majority of the 250 publications on BSE fall in only two software engineering areas: \textit{software engineering professional practice} and \textit{software engineering management}. While BSE concepts are related to human values, they include a wider range of human aspects (e.g., cognitive, behavioral, and social) in software engineering. Lenberg et al.'s work also mainly focused on providing a definition for BSE and identifying BSE concepts.

All previous review studies have been vital to showing the importance of values in software by focusing on an important goal (e.g., the relevance of the software engineering literature to values) or targeting a specific aspect of software development (e.g., VBSE or VSD techniques). Our survey attempts to build on this body of knowledge and provides a comprehensive view of operationalizing human values in software engineering by identifying and classifying solutions that can be applied in any aspect of software development. We also provide a taxonomy of the solutions and discuss what values are operationalized.
\section{Research Method}\label{sec:researchmethod}
In the following sections, we define our research question and outline the scope of our survey and the execution of the survey.
\subsection{Research Question}
Researchers, organizations, and practitioners have been trying to incorporate human values in technology development (e.g., medical devices) \cite{manders2011values}. Our main goal in this survey is to identify and classify solutions that help operationalize human values in software. By solutions, we mean any techniques, approaches, practices, frameworks, tools, etc., that can be used in or support the process of operationalizing human values in software engineering.
Hence, we formulate the following research question (RQ):

\textit{\textbf{RQ.} What solutions do exist to enable or support operationalizing human values in software engineering?}




\subsection{Survey Scope}\label{sec:paperscope}
We apply the following inclusion/exclusion criteria to collect and select primary studies for our survey.
\begin{itemize}
\item The paper should propose a solution that can be applied to one or more software engineering phases/aspects. For example, the papers that offer solutions to address values in non-software systems (e.g., car design) are excluded.
\item The papers that test a set of hypotheses on human values (e.g., investigating the relationship between human values and e-learning adoption \cite{mehta2019influence}) without proposing any solutions to integrate values are excluded.

\item If we find papers that present the same solution but appear in diverse venues (e.g., a conference paper and its extension in a journal), we include the most mature paper.
\item As described in Section \ref{sec:Background}, \textit{security} and \textit{privacy} are two human values that appear in many human values models (e.g., the Schwartz theory of basic values). We exclude the papers that provide technical solutions to improve \textit{security} or \textit{privacy} in the software for two reasons. First, such technical solutions are out of the scope of our study. Second, \textit{security} and \textit{privacy} have been extensively studied in the past decade in the software and systems engineering communities, and a substantial number of reviews exist on these two concepts.
\item \textit{Fairness} can be considered a human value. There are many papers in the AI and machine learning community that attempt to detect and address \textit{fairness} issues in machine learning {algorithms} and {models} \cite{zhang2020machine,mehrabi2019survey}. We exclude such papers if they do not investigate \textit{fairness} from the software engineering perspective, as this is not within our survey's scope. 
\item As we described in Section \ref{sec:Definitions}, human values and the concepts such as motivation, ethics, and emotion are intertwined. Therefore, if a paper proposes a solution to operationalize these concepts in software engineering with human values-related examples and discussions, we will include such papers.
\end{itemize}
\subsection{Paper Collection}\label{sec:papercolection}
There are two different techniques to identify the primary sources for literature review studies \cite{jalali2012systematic}. In the first technique, which is common in the software engineering community, search strings are developed and then executed on different digital libraries (e.g., ACM Digital Library) \cite{kitchenham2007guidelines}. The second one is more common in the information systems community and starts with identifying a pool of initial papers, followed by the backward snowballing technique \cite{webster2002analyzing}. Jalali and Wohlin \cite{jalali2012systematic} applied both techniques on Agile practices in Global Software Engineering (GSD) and realized that although these techniques led to the identification of different sets of studies, no significant differences were observed in the findings.

\begin{figure*}
    \centering
    \includegraphics[width=0.95\textwidth]{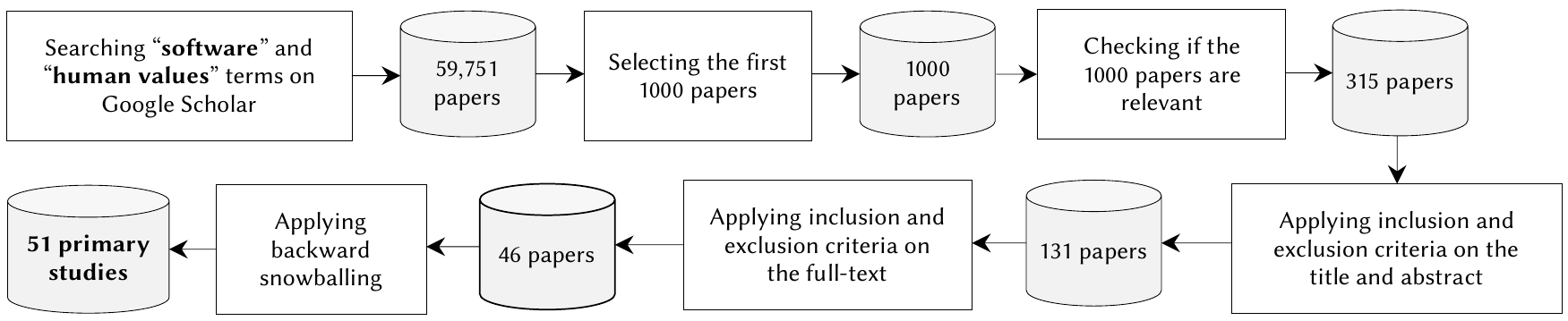}
    \caption{The process of paper collection. }
    \label{fig:papercollection}
\end{figure*}

Human values have been researched in many domains across different research areas. In Section \ref{sec:Background}, we discussed that there is no consensus on what human values are, and there are many values models that cover a different number of human values with various terminologies. Further to this, there is no established theory on human values within the software engineering community \cite{Perera2020}. Due to these limitations, it was not possible for us to build a search string that covers all human values and execute it on different digital libraries. Hence, we decided to follow the approach proposed by Webster and Watson in the information systems community, which includes the following two steps \cite{webster2002analyzing}. Figure \ref{fig:papercollection} shows our paper collection process.
\subsubsection{Initial Set of Papers}
This step included four phases. In the first phase, we searched on Google Scholar search engine with two general terms: ``human values'' and ``software''. We looked at only the title of the first 1000 out of 59,751 papers returned by Google Scholar and excluded 685 papers as they did not meet some of our inclusion or exclusion criteria. Next, the first author read the abstract of all these 315 papers and chose 131 papers that had the potential to be included in this survey. As shown in Figure \ref{fig:papercollection}, the inclusion and exclusion criteria outlined in Section \ref{sec:paperscope} were applied on the abstract of the 315 papers to select these 131. We maintained an Excel spreadsheet file to record which papers were included or excluded in these two first steps and the rationale behind our decisions. We shared the file with four of the authors to receive their feedback. Next, the 131 papers were distributed among five of the authors. They were asked to read the full text of the papers and determine which of their assigned papers proposed solutions for operationalizing values in software. They had to record the reason for including or excluding a paper in the Excel spreadsheet file for further discussions. Finally, 46 papers met all the inclusion and exclusion criteria and were included in our survey.

\subsubsection {Backward Snowballing} 
We used the backward snowballing technique \cite{wohlin2014guidelines} to minimize the risk of missing pertinent studies. Following the guidelines proposed by Wohlin \cite{wohlin2014guidelines}, the first author conducted the backward snowballing in several iterations until no new papers were found. The first author checked the references of all 46 papers and employed the inclusion and exclusion criteria discussed in Section \ref{sec:paperscope}. Similar to the \textit{``Initial Set of Papers''} step, the rationale behind excluding a paper was recorded in the Excel spreadsheet file and shared with four of the authors for seeking their comments. This step added 5 papers to the pool of primary studies. Table \ref{tbl:papersID} shows all 51 primary studies that are studied in this survey. It is worth noting that we did not conduct the forward snowballing. It is because the significant time required to conduct a systematic literature study often forces the majority of researchers to limit their search procedures (See Section \ref{limiations}) \cite{jalali2012systematic}.
\subsection{Data Extraction}
The first author created another Excel spreadsheet file shared with five of the authors to extract the detailed contents from the 51 primary studies. The 51 primary studies were distributed between the first six authors to extract the relevant information: the first author (27 primary studies), the second one (9 primary studies), the third one (5 primary studies), the fourth one (5 primary studies), the fifth one (4 primary studies), and the sixth one (1 primary study). This allocation was based on the time availability of the authors. We collected the following data items from each primary study. Table \ref{tbl:DataExtraction} shows these data items.
\begin{itemize}
  \item \textbf{D1}-\textbf{D9}: These data items were collected to provide comprehensive demographic information on the primary studies. We obtained the citations (D3) of each primary study from Google Scholar on 24 March 2021.
  \item \textbf{D10}: We extracted how a solution proposed in a given primary study for operationalizing values is evaluated (D10).
  \item \textbf{D11}: Our study recorded if a solution operationalizes a specific human value or is designed to operationalize any human values.
  \item \textbf{D12}-\textbf{D13}: Some researchers prefer to give a name to their proposed solutions. In such a case, we recorded the name of the solution (D12). Some of the solutions are supported by tools. Hence, we recorded if a solution is supported by a tool and collected the tool's name, provided that it was mentioned (D13). 
  \item \textbf{D14}-\textbf{D17}: We wrote a critical summary of how a solution supports or enables operationalizing values (D14). We also recorded the problems (D15) solved by the solution and the benefits (D16) of the solution. Finally, we collected the possible limitations (D17) of the solution.
\end{itemize}
\begin{table*}[]
\caption{Data extraction form} 
\centering
\footnotesize
\label{tbl:DataExtraction}
\begin{tabular}{lll|lll}
\toprule
\textbf{\#}  & \textbf{Data Item}              & \textbf{Used in}    & \textbf{\#}  & \textbf{Data Item}              & \textbf{Used in}   \\ \toprule
D1  & Title                  & Section \ref{sec:venues} & D10  & Research Type          & Section \ref{sec:researchtype}\\
D2  & Author(s)              & Section \ref{sec:authorsafflications} & 
D11 & Operationalized Value  & Section \ref{sec:taxonomy} \& Section \ref{sec:targtedvalues}\\
D3  & Citation               & Section \ref{sec:citations} & D12 & Solution Name (if any) & Section \ref{sec:taxonomy}\\
D4  & H5-index               & Section \ref{sec:citations} & D13 & Tool Name (if any) & Section \ref{sec:taxonomy} \& Section \ref{sec:toolsupport}\\
D5  & Affiliation            & Section \ref{sec:authorsafflications} & D14 & Solution Summary       & Section \ref{sec:taxonomy} \\
D6  & Country                & Section \ref{sec:authorsafflications} & D15 & Problems       & Section \ref{sec:taxonomy} \\
D7  & Year                   & Section \ref{sec:years} & D16 & Benefits                & Section \ref{sec:taxonomy} \\
D8  & Venue                  & Section \ref{sec:venues} & D17 & Limitations               & Section \ref{sec:taxonomy} \\
D9  & Publication Type       & Section \ref{sec:venues} & - & -            & -\\
\bottomrule
\end{tabular}
\end{table*}
\subsection{Data Analysis}\label{sec:datanalysis}
Data items D1 to D9 and D11 to D13 were analyzed using descriptive statistics. We used the taxonomy proposed by Wieringa et al. \cite{wieringa2006requirements} to classify the research type used to evaluate a solution (e.g., a technique, tool, etc.). Wieringa et al. suggest that the evaluation of a solution (in the requirements engineering community) can be done through the following six research types:  ``validation research'',  ``evaluation research'',  ``solution proposal'',  ``philosophical paper'',  ``opinion paper'', and  ``experience report'' \cite{wieringa2006requirements}. Note that Wieringa et al.'s taxonomy has been widely used in many review papers in the software engineering community to classify research types. For example, Engström and Runeson \cite{engstrom2011software} used it to classify papers on software product line testing, and Jalali and Wohlin \cite{jalali2012systematic} used it in a review on global software engineering.

The data collected from data items D14 to D17 were analyzed using the open coding procedure \cite{seaman1999qualitative} to build a taxonomy of solutions for operationalizing values in software. The first author performed open coding iteratively in parallel with data extraction and labeled the data. First, the first author read all the extracted data to become familiar with the extracted data. He constantly contacted the authors involved in data extraction if any ambiguity or missing information (e,g., a description of a solution was not understandable) was found in the data. Next, he coded each study and shared the codes with the author responsible for reading and extracting the data from the given study to seek their feedback on the identified codes. The first author updated the codes based on feedback and comments from the corresponding authors. In the next step, codes identified in one study were compared with those that emerged from other studies. The next step iteratively classified these emergent labels to build the taxonomy. In the last step, the taxonomy was shared with other authors to seek their feedback. Any disagreements between the authors were solved by organizing several face-to-face and Zoom meetings. The final version of the taxonomy was agreed upon by all the authors.

\section{Primary Studies Demographics}\label{sec:demographics}




\Cref{tbl:papersID} shows the 51 primary studies selected for analysis in this survey. We summarize the demographics of these primary studies in the followings sections.

\begin{table*}[]
\caption{All 51 primary studies in this study (J: Journal; C: Conference; BC: Book Chapter; W: Workshop)}
\label{tbl:papersID}
\resizebox{\textwidth}{!}{%
{\renewcommand{\arraystretch}{1.1}
\begin{tabular}{cclccc}
\toprule
\textbf{Ref} & \textbf{ID} & \textbf{Title} & \textbf{V. Type}  & \textbf{Year}  & \textbf{Citation}                                                                                             \\ \toprule
\cite{thew2018value}  & P1  & Value-based Requirements Engineering: Method and Experience  & J & 2018 & 42                                                                                         \\
 \cite{winter2018measuring} & P2   & Measuring Human Values in Software Engineering    & C & 2018 &25                                                                                                                                                                                            \\
\cite{ferrario2016values} & P3   & Values-First SE: Research Principles in Practice      & C & 2016 &43                                                                                                                                                                                          \\
\cite{zdravkovic2015capturing} & P4   & Capturing Consumer Preferences as Requirements for Software Product Lines                     & J & 2015 &20                                                                                                                                                 \\
\cite{pereira2015value} & P5   & A Value-oriented and Culturally Informed Approach to the Design of Interactive Systems            & J & 2015 &63                                                                                                                                             \\
\cite{barn2015role} & P6   & On the Role of Value Sensitive Concerns in Software Engineering Practice       & C & 2015 &17                                                                                                                                                                \\
\cite{aldewereld2015design} & P7   & Design for Values in Software Development       & BC & 2015 &25                                                                                                                                                                                                \\
\cite{pereira2014value} & P8  & Value Pie: A Culturally Informed Conceptual Scheme for Understanding Values in Design       & C & 2014 &23                                                                                                                                                    \\
\cite{anthonysamy2012method} & P9  & A Method for Analysing Traceability between Privacy Policies and Privacy Controls of Online Social Networks         & C & 2014 &15                                                                                                                       \\
\cite{koch2013approximate} & P10  & 
How to Approximate Users’ Values while Preserving Privacy: Experiences with Using Attitudes towards Work ... 
& J & 2013 &10                                                                                   
\\
\cite{sjokvist2019eliciting} & P11  & Eliciting Human Values by Applying Design Thinking Techniques in Systems Engineering              & J & 2019 & 2   \\

\cite{yoo2013value} & P12  & A Value Sensitive Action-Reflection Model: Evolving a Co-Design Space with Stakeholder and Designer Prompts         & C & 2013 &99                                                                                                                           \\
\cite{iversen2012values} & P13  & Values-led Participatory Design          & J & 2012 &146                                                                                                                                                                                                      \\
\cite{colomo2011using} & P14  & Using the Affect Grid to Measure Emotions in Software Requirements Engineering          & J & 2011 &67                                                                                                                                                       \\
\cite{detweiler2010principles} & P15  & Principles for Value-Sensitive Agent-Oriented Software Engineering                     & W & 2010 &10                                                                                                                                                        \\
\cite{proynova2011investigating} & P16  & Investigating the Influence of Personal Values on Requirements for Health Care Information Systems           & W & 2011 &20                                                                                                                                     \\
\cite{li2009bridge} & P17  & Bridge the Gap between Software Test Process and Business Value: A Case Study                & C & 2009 &24                                                                                                                                                   \\
\cite{perera2020continual} & P18  & Continual Human Value Analysis in Software Development: A Goal Model Based Approach              & C & 2020 & 1   \\

\cite{nathan2008envisioning} & P19  & Envisioning Systemic Effects on Persons and Society Throughout Interactive System Design   & C & 2008 &104                                                              \\
\cite{ramos2005requirements} & P20  & Requirements Engineering for Organizational Transformation        & J & 2005 &55                                                                                                                                                                             \\
\cite{miller2005agile} & P21  & Agile Software Development: Human Values and Culture     & J & 2005 &35                                                                                                                                                                              \\
\cite{flanagan2005values} & P22  & Values at Play: Design Trade-offs in Socially-Oriented Game Design                                                      & C & 2005 &189                                                                                \\
\cite{cockton2005development} & P23  & A Development Framework for Value-Centred Design                            & C & 2005 &137                                                                             \\
\cite{mussbacher2020there} & P24  & Is There a Need to Address Human Values in Domain Modelling?              & W & 2020 & 0   \\
\cite{surian2011recommending} & P25  & Recommending People in Developers' Collaboration Network                 & C & 2011 &76                                                                                \\
\cite{lee2014customer} & P26  & Customer Requirements Validation Method Based on Mental Models                 & C & 2014 &7                                                                          \\
 \cite{schuler2013rule} & P27  & Rule-Based Generation of Mobile User Interfaces            & C & 2013 &5                                                                                             \\
 \cite{amreen2019developer} & P28  & Developer Reputation Estimator (DRE)     & C & 2019 &0                                                                                                              \\
\cite{rathnayake2019framework} & P29  & A Framework for Adaptive User Interface Generation based on User Behavioural Patterns            & C & 2019 &1                                                        \\
\cite{romero2019adapting} & P30  & Adapting Scrum Methodology to Develop Accessible Web Sites               & C & 2019 &1                                                                                \\
\cite{barivsic2017requirements} & P31  & A Requirements Engineering Approach for Usability-driven DSL Development     & C & 2017 &5                                                                            \\
\cite{albarghouthi2019fairness} & P32  & Fairness-Aware Programming            & C & 2019 &14                                                                                                                  \\
\cite{ying2016earec} & P33  & EARec: Leveraging Expertise and Authority for Pull-request Reviewer Recommendation in GitHub                   & W & 2016 & 15                                           \\
\cite{pellegrini2019prioritize} & P34  & How to Prioritize Accessibility in Agile Projects                                    & C & 2020 & 2                                                                   \\
\cite{de2020moderating} & P35  & On Moderating Software Crowdsourcing Challenges             & J & 2020 & 3                                                                                            \\
\cite{curumsing2019emotion} & P36  & Emotion-oriented Requirements Engineering: A Case Study in Developing a Smart Home System for the Elderly                & J & 2019& 13                               \\
\cite{harbers2015embedding}  & P37  & Embedding Stakeholder Values in the Requirements Engineering Process          & C & 2015 & 21                                                                           \\
\cite{pedell2015don} & P38  & Don’t Leave Me Untouched: Considering Emotions in Personal Alarm Use and Development              & BC & 2015 & 14                                                       \\
 \cite{kheirandish2019huvalue} & P39  & HuValue: a Tool to Support Design Students in Considering Human Values in their Design  & J & 2019 & 6                                                                \\
\cite{giorgini2006requirements} & P40  & Requirements Engineering for Trust Management: Model, Methodology, and Reasoning          &J & 2006 & 106                                                               \\
\cite{uddin2008umltrust} & P41  & UMLtrust: Towards Developing Trust-Aware Software        & C & 2008 & 21                                                                                                         \\
\cite{doerr2007built} & P42  & Built-in User Satisfaction – Feature Appraisal and Prioritization with AMUSE             & C & 2007 & 28                                                                \\
\cite{mchugh2011agile} & P43  & Agile Practices: The Impact on Trust in Software Project Teams             & J & 2012 & 148                                                                              \\
\cite{aydemir2018roadmap} & P44  & A Roadmap for Ethics-Aware Software Engineering                                      & W & 2018 & 15                                                                   \\
\cite{lopez2019talking} & P45  & Talking about Security with Professional Developers                         & W & 2019 & 5                                                                             \\ 
\cite{galhotra2017fairness} & P46  & Fairness Testing: Testing Software for Discrimination              & C & 2017 & 148                                                                                    \\ 
\cite{sutcliffe2006pc} & P47  & PC-RE: A Method for Personal and Contextual Requirements Engineering with some Experience              & J & 2006 &    104\\
\cite{liaskos2011representing} & P48  & Representing and Reasoning about Preferences in Requirements Engineering              & J & 2011 & 127   \\
\cite{tramer2017fairtest} & P49  & FairTest: Discovering Unwarranted Associations in Data-Driven Applications              & C & 2017 & 72   \\
\cite{sindre2005eliciting} & P50  & Eliciting Security Requirements with Misuse Cases              & J & 2008 & 1368   \\
\cite{mougouei2020engineering} & P51  & Engineering Human Values in Software through Value Programming              & W & 2020 & 2   \\
\bottomrule
\end{tabular}
}\quad
}
\end{table*}

\subsection{Venues}\label{sec:venues}
\begin{figure}
    \centering
    \includegraphics[width=0.35\textwidth]{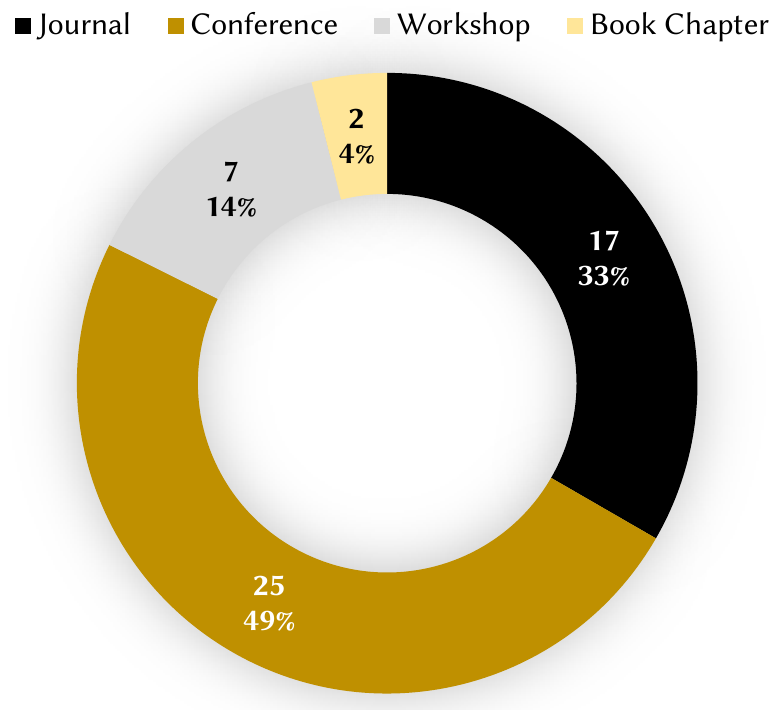}
    \caption{Primary studies over types of venues.}
    \label{fig:papervenue}
\end{figure}

\Cref{fig:papervenue} shows that conferences are the dominant venues to publish research on operationalizing values in software. 25 primary studies (49\%) were published in conferences, 17 (33\%) primary studies in journals, and the rest were workshop papers (7 primary studies, 14\%) and book chapters (2 primary studies, 4\%). Table \ref{tbl:popularvenues} reveals that the 51 primary studies come from 42 distinct venues, in which \textit{``Requirements Engineering Journal''} with 5 primary studies is the most popular one, followed by \textit{``ACM Conference on Human Factors in Computing Systems''} (4 primary studies). There are two venues with two primary studies each and 38 venues with only one paper each. This indicates that research on human values does not have exclusive venues and attracts a wide range of researchers from computer sciences, software engineering, and human-computer interaction.    
\begin{table}[]
\caption{The publication venue of the 51 primary studies.}
\label{tbl:popularvenues}
\footnotesize
\begin{tabular}{lcc}
\toprule
\textbf{Venue Name}                                                                          & \textbf{\#} & \textbf{\%} \\ \toprule
Requirements Engineering Journal                                         & 5          & 9.8         \\
ACM Conference on Human Factors in Computing Systems                                                                & 4           & 7.8         \\
International Requirements Engineering Conference                                             & 2           & 3.9          \\
International Conference on Software Engineering                                            & 2           & 3.9         \\

Others                                                                                       & 38          & 74.5        \\ \bottomrule
\end{tabular}%
\end{table}
\subsection{Affiliations and Countries}\label{sec:authorsafflications}

The authors of the 51 primary studies come from 22 countries. We found that researchers from the USA (13 primary studies, 863 citations), UK (9 primary studies, 514 citations), and Australia (7 primary studies, 193 citations) have contributed more to this research area than others (See Table \ref{tbl:researchimpact}). In total, 78 institutes published in this area. The Delft University of Technology had 4 primary studies, and researchers from Washington University and Lancaster University published three primary studies each (See \ref{tbl:researchimpact}). The vast majority of them (71 institutes) had only one primary study. Note that we did not find any authors with more than two papers on operationalizing human values in software.
\begin{table*}[]
\caption{The analysis of top countries and institutes}
\centering
\label{tbl:researchimpact}
\footnotesize
\begin{tabular}{lcc|lcc}
\toprule

\multicolumn{3}{c|}{\textbf{Top 3 Countries}} & \multicolumn{3}{c}{\textbf{Top 3 Affiliations}} \\ \hline
Country         & Paper & Citation & Affiliation    & Paper    & Citation   \\ \hline
USA             &    13   &    863      & Delft University of Technology    &    4      &        155    \\
UK              &   9    &      514    & University of Washington    &    3      & 213           \\
Australia &     7  &    193      & Lancaster University    &     3     &       83     \\  \bottomrule
\end{tabular}
\end{table*}
\subsection{Years of Publication}\label{sec:years}
\begin{figure*}
    \centering
    \includegraphics[width=0.99\textwidth]{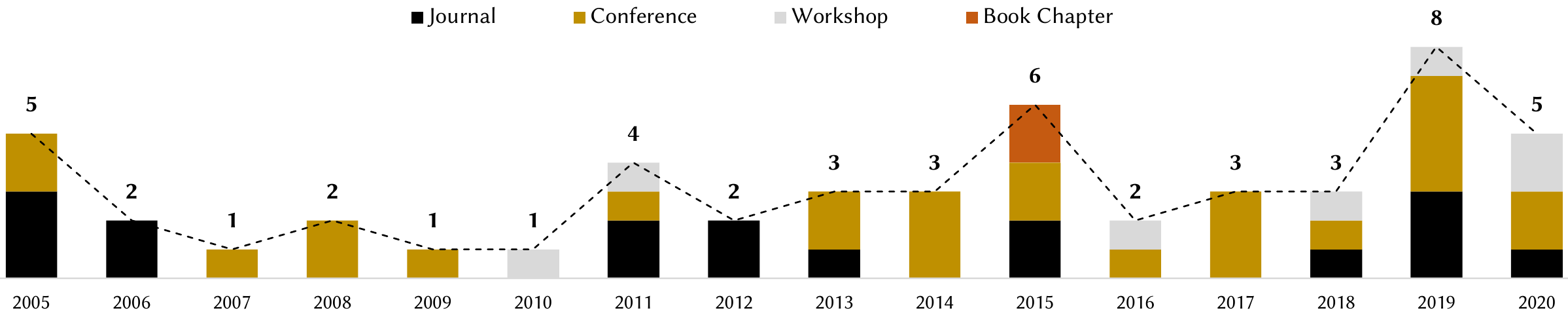}
    \caption{Primary studies in each year and primary studies in each venue type.}
    \label{fig:paperdistribution}
\end{figure*}

\Cref{fig:paperdistribution} shows the number of primary studies published from 2005 to 2020. The average number of primary studies in this research area is 3 studies per year with 3.9 papers in the last ten years, where 2019 peaked (8 primary studies), followed by 2015 (6 primary studies). This indicates that the interest in the research on human values in software engineering seems to remain more or less constant since 2011.
\subsection{Citations}\label{sec:citations}
\begin{figure*}
    \centering
    \includegraphics[width=0.70\textwidth]{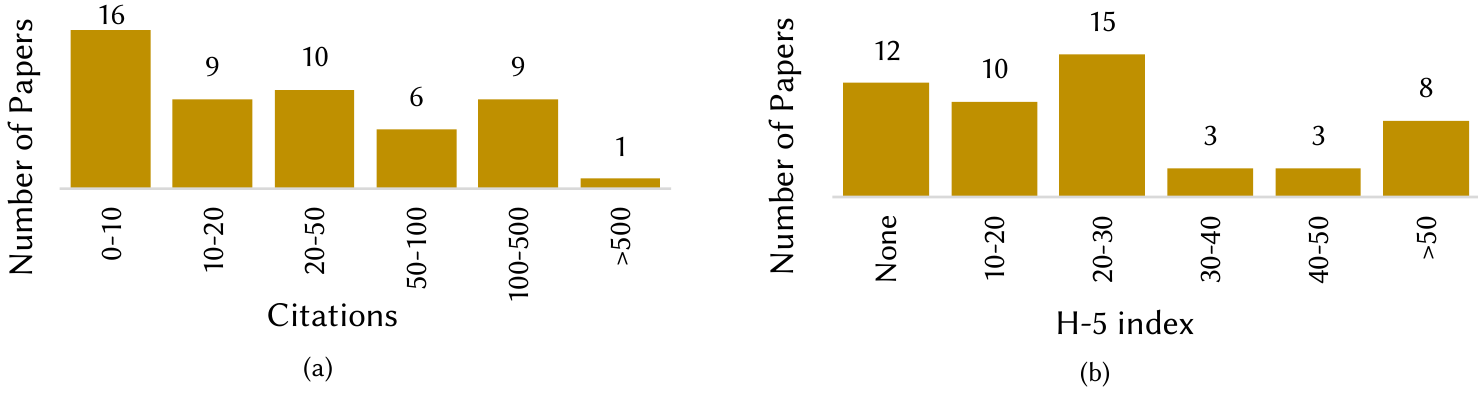}
    \caption{Quality assessment of primary studies with the number of citations and H5-index of their venues.}
    \label{fig:citation}
\end{figure*}

Citations and venues of a paper can partially show the quality of the research paper \cite{aksnes2003characteristics}. We obtained the citation counts of the 51 primary studies from Google Scholar on 24 March 2021. \Cref{tbl:papersID} indicates that the number of citations ranges from 0 to 1368, with a high average of 68.6. Figure \ref{fig:citation}.(a) shows that 50\% of the primary studies have more than 20 citations.
We also used the Google Scholar H5-index\footnote{https://scholar.google.com/intl/en/scholar/metrics.html} of the primary studies' venues as an indicator to judge the quality of the primary studies. A higher H5-index implies that more quality papers appear in the venue. As shown in Figure \ref{fig:citation}.(b), there are 12 primary studies published in the venues whose H5-index were not available (e.g., new conferences, workshops) and were labeled as ``None''. The average H5-index for the remaining venues is 35. Figure \ref{fig:citation}.(b) shows that most of the primary studies (29 out of 51, 56.8\%) appeared in venues with an H5-index of more than 20. These results can (partially) show that research on human values is being published in quality venues.

\subsection{Research Types} \label{sec:researchtype}
As we described in Section \ref{sec:datanalysis}, we used the taxonomy provided by Wieringa et al. \cite{wieringa2006requirements} to evaluate the solutions proposed in the primary studies. As illustrated in Table \ref{tbl:evalutionType}, most of the primary studies (29 out of 51 primary studies, 56.8\%) are classified as ``solution proposals''. This group of primary studies introduces a solution and discusses its effectiveness, but usually without a solid and sufficient validation. Such primary studies examined the usefulness and actionability of their proposed solution by a small example application, sound argument, case studies, or experiments. We found 13 primary studies (P1, P2, P3, P5, P9, P14, P36, P37, P38, P39, P42, P46, P50) studying attributes of a solution that was not implemented in the industry (i.e. ``validation paper''). We classified 6 primary studies (P13, P16, P17, P22, P35, P43) as ``evaluation research''. These primary studies investigated the (positive and negative) consequences of a solution that was implemented in practice. 

Only one primary study is assigned to each of the ``philosophical paper'' (P15), ``opinion paper'' (P34), and ``experience report'' (P30) categories. P15 is classified as a philosophical paper because it proposes new notations and concepts for Tropos to support values in the software development process. In P34, the authors report their personal opinions about how the responsibilities of Scrum Master, Product Owner, and Development Team should be adapted to ensure accessibility in Agile projects. Based on an experience report, the authors in P30 suggest how to adjust Scrum (e.g., adding new tasks to Scrum) to meet accessibility requirements in a software project. 

Given the low portion of the papers with industrial evidence, a call for conducting more validation and evaluation research is demanded. Such research improves the practical applicability of solutions proposed to operationalize human values and encourages software development organizations to adopt those solutions in practice.
\begin{table*}[]
\caption{Number of the primary studies in each research type}
\label{tbl:evalutionType}
\resizebox{\textwidth}{!}{%
{\renewcommand{\arraystretch}{1.04}
\begin{tabular}{lll}
\toprule
\textbf{Research Type}       & \textbf{Included Primary Papers }                                                                                                                                                                       & \textbf{N}  \\ \toprule
Validation Research & P1, P2, P3, P5, P9, P14, P36, P37, P38, P39, P42, P46, P50                                                                                                                             & \textbf{13} \\
Evaluation Research & P13, P16, P17, P22, P35, P43                                                                                                                                                           & \textbf{6}  \\
Solution Proposal   & \begin{tabular}[c]{@{}l@{}}P4, P6, P7, P8, P10, P11, P12, P18, P19, P20, P21, P23, P24, P25, P26, \\ P27, P28, P29, P31, P32, P33, P40, P41, P44, P45, P47, P48, P49, P51\end{tabular} & \textbf{29} \\
Philosophical Paper & P15                                                                                                                                                                                    & \textbf{1}  \\
Opinion Paper       & P34                                                                                                                                                                                    & \textbf{1}  \\
Experience Report   & P30                                                                                                                                                                                    &\textbf{1}  \\ \bottomrule
\end{tabular}
}\quad
}
\end{table*}


\section{Solutions for Operationalizing Human Values in Software}\label{sec:taxonomy}
\begin{figure*}
   
   \centering
    \includegraphics[width=0.95\textwidth]{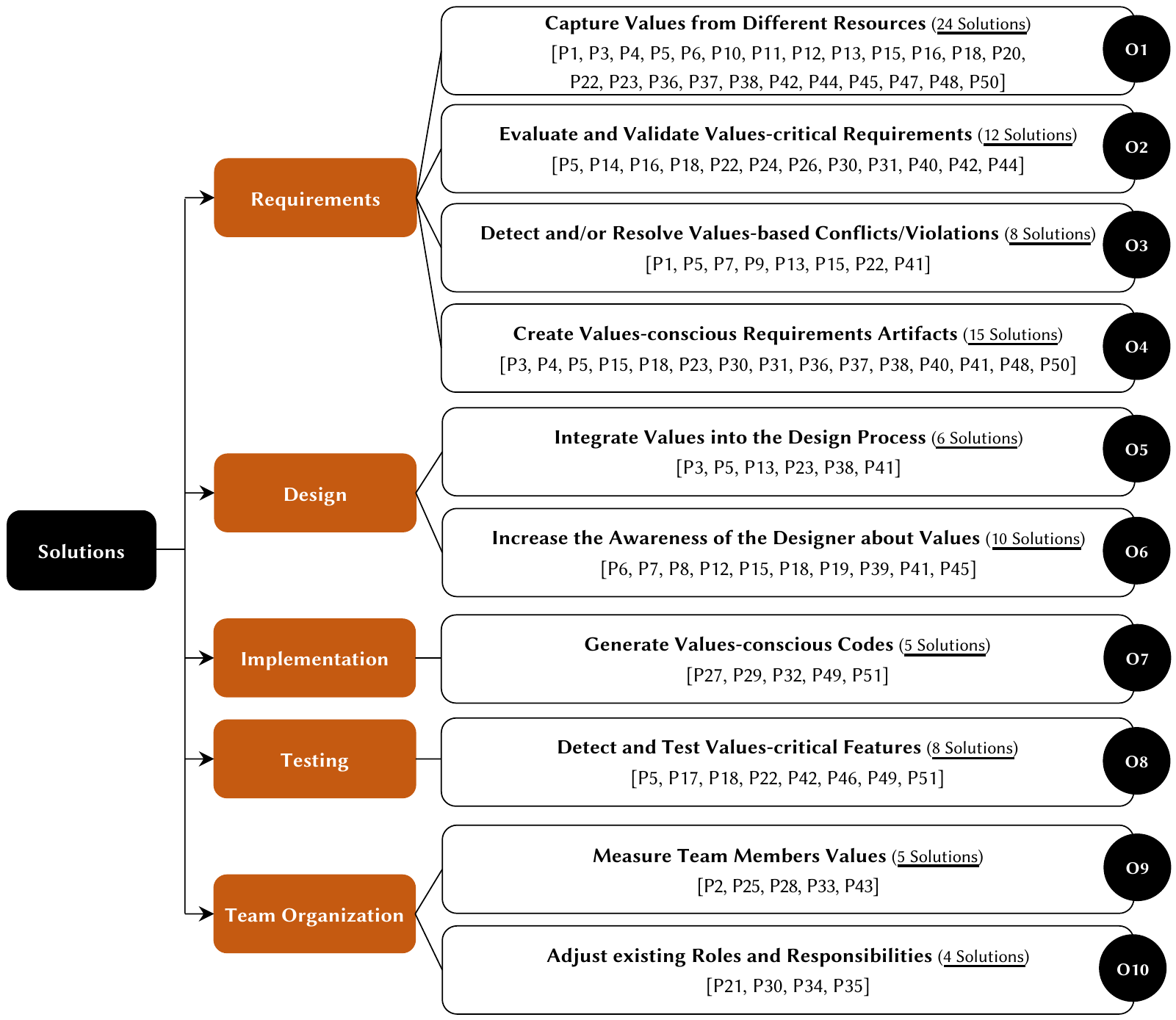}
    \caption{A taxonomy of solutions for operationalizing values.}
     \label{fig:classifcation_new}
\end{figure*}

We identified 51 solutions from the 51 primary studies (each of the primary studies proposes a solution to operationalize human values in software). As we described in Section \ref{sec:datanalysis}, we applied the open coding procedure (on data items D14 to D17) to analyze these 51 solutions, leading to a taxonomy. Figure \ref{fig:classifcation_new} shows the taxonomy of the 51 solutions. The identified solutions can be applied at the highest level in the following five areas: \textit{requirements}, \textit{design}, \textit{implementation}, \textit{testing}, and \textit{team organization}. In the next level, we further classified the solutions into 10 categories (\textbf{O1} to \textbf{O10} in Figure \ref{fig:classifcation_new}). The 10 categories are not mutually exclusive, as there might be solutions that fall in more than one category. 
Table \ref{tbl:AppToolsValues} in the Appendix provides a comprehensive overview of all these solutions. 
For brevity, a small subset of the solutions in each category as examples is elaborated in the following sections.




\subsection{Requirements} 

\subsubsection{\textbf{Capture Values from Different Resources}}\label{sec:CaptureValues} According to the definition provided in the Introduction section, the first step of operationalizing values in software is to identify values. However, understanding and eliciting human values are not a non-trivial task. As highlighted in Table \ref{tbl:valuesdiff}, this difficulty stems from three main factors:
\begin{itemize}
    \item \textbf{Values' characteristics}. Values are tacit knowledge and subjective concepts, and cannot be (easily) understood outside of their social and cultural context.
    \item \textbf{Stakeholders' characteristics}. It is not easy to understand and obtain stakeholder values as individuals and organizations seldom immediately expose such information.
    \item \textbf{Lack of practical solutions}. The software development methods widely used in the software industry (e.g., Agile methods) lack any practical values-based guidelines to elicit human values.
\end{itemize}
\begin{table*}[]
\caption{Factors that make values elicitation challenging}
\centering
\label{tbl:valuesdiff}

\begin{tabular}{lll}

\toprule
\textbf{Factors}                      &                                                    & \textbf{Included Primary Studies} \\ \toprule
Values' characteristics       & Values are tacit knowledge                         &            P1, P26, P34, P36              \\ \cline{2-3} 
                             & Values are subjective concepts                     &      P2, P36, P42                    \\ \cline{2-3} 
                             & Values depend on their social and cultural context &   P8, P12, P13, P23, P42                       \\ \hline
Stakeholders' characteristics &                                                    &       P1, P12, P42, P45                   \\ \hline
Lack of practical solutions  &                                                   &            P3, P6, P7, P36, P39               \\ \bottomrule
\end{tabular}%
\end{table*}
Hence, it would be difficult for the software development team to extract and document human values. A large number of primary studies (P1, P3, P4, P5, P6, P10, P11, P12, P13, P15, P16, P18, P20, P22, P23, P36, P37, P38, P42, P44, P45, P47, P48, P50) develop solutions to elicit values from the relevant resources and help stakeholders communicate and discuss values explicitly during the requirements engineering process. 

In Table \ref{tbl:sourcesofvalues}, we further classify these primary studies based on the sources that they investigate to elicit values. As shown in Table \ref{tbl:sourcesofvalues}, 14 primary studies attempt to identify values from any relevant stakeholders. Stakeholders are defined as anyone who is directly or indirectly influenced by a software-intensive system \cite{Sommerville2016}. Hence, stakeholders can be any combination of end-users, development teams, customers, and organizations. Six primary studies only use end-users, and one primary study only uses development team members for this purpose. Other sources for values elicitation are software development artifacts (e.g., requirements documents), existing systems, literature, and prototypes. Table \ref{tbl:instruments} indicates that interviews and surveys are the main instruments to elicit values, followed by co-design workshops. Three primary studies (P1, P10, P11) use observations to extract values from stakeholders. Below are some good examples to illustrate how values can be elicited from different resources using the instruments presented in Table \ref{tbl:instruments}.
\begin{table*}[]
\caption{Sources that are investigated to elicit values.} 
\centering
\label{tbl:sourcesofvalues}
\footnotesize
\begin{tabular}{l|l|c}
\toprule
\textbf{Sources of Values}                                            & \textbf{Primary Studies}                                                     & \textbf{\#} \\ \toprule
All Relevant Stakeholders & P1, P3, P5, P6, P11, P12, P13, P15, P18, P20, P22, P37, P44, P50 &   14 \\ \midrule
Only End-Users                                                    & P4, P10, P16, P36, P38, P42                            &   6 \\ \midrule
Software Development Artifacts                                                 & P1, P22, P23                                           &  3   \\ \midrule
Existing Systems                                             & P5, P22                                                     &  2   \\ \midrule

Only Development Team                                             & P45                                                    & 1  \\ \midrule

Literature                                                   & P22                                                         &   1  \\ \midrule

Prototypes                                                   & P47                                                         &   1 \\  \bottomrule
\end{tabular}%
\end{table*}
\begin{table*}[]
\caption{Instruments to elicit values}
\centering
\label{tbl:instruments}
\footnotesize
\begin{tabular}{l|l|c}
\toprule
\textbf{Instruments}         & \textbf{Primary Studies (Handouts) }                                                                                                                                                                            & \textbf{\# }\\ \toprule
Interview           & P1, P3, P4, P5, P10, P11, P15, P36, P38, P47                                                                                                                                                 & 10 \\ \midrule
Survey              & P3, P4, P5, P10, P16, P18, P36, P38, P42                                                                                                                                                & 9 \\ \midrule
Co-design Workshops & \begin{tabular}[c]{@{}l@{}}P1, P5 (Existing Systems), P6 (Prompts), P12 (Stakeholder Prompt,\\ Designer Prompt), P13 (Inspiration Card Workshop), P37, P45 (Prompts \\and Values Cards with definitions)\end{tabular} & 7  \\ \midrule
Observation         & P1, P10, P11                                                                                                                                                                                  & 3  \\ \midrule
Feedback            & P3                                                                                                                                                                                            & 1  \\ \midrule
Focus Group         & P47                                                                                                                                                                                           & 1  \\ \midrule
Wizard of OZ Method & P47                                                                                                                                                                                           & 1  \\ \midrule
CORE Method      & P47                                                                                                                                                                                           & 1  \\ \midrule
Applying Guidelines & P50 & 1 \\ \bottomrule
\end{tabular}%
\end{table*}

Thew et al. (P1) propose \textit{Value-based Requirements Engineering} (\textit{VBRE}) to make values explicit in the requirements engineering process. The \textit{VBRE} process provides a set of step-by-step guidelines that advise and assist the analyst in obtaining values from the requirements engineering artifacts, documents, and interviews. In this process, the analyst is provided with a paper-based taxonomy of values, which is supported by a website. The proposed taxonomy may not be easily understandable by the analyst. Hence, the supporting website provides detailed information about each item (e.g., value) in the taxonomy. This includes a list of representative interview questions and scenarios to support the analyst, particularly the novice analyst, identify and capture values.

\textit{Emotion-oriented Requirements Engineering} technique (P36) uses a set of notations to model emotional goals (``how it feels''), functional goals (``what it does''), and quality (non-functional) goals (``how it does''). The technique first extracts ``emotional threats'' (users' pain points) from the relevant stakeholders using an emotion-oriented interview and survey. Then each extracted threat will be translated into emotional, functional, and quality goals. For example, the ``insecure'' threat can be converted into ``safe'' as an emotional goal, ``responsiveness'' as a quality goal, and ``anomaly detection'' as a functional goal. Then, all functional goals should be further decomposed into sub-functional goals. Any emotional goal must be linked to one or more functional goals.

Koch et al. (P10) argue that users are reluctant to talk about their personal values because they are concerned about their privacy and are afraid of their personal (secret) information being disclosed. Hence, directly asking users about their values is not appropriate. Koch et al. suggest a technique for predicting the values of end-users. This prediction is achieved by determining the preferences of end-users for tasks that they do in a work domain.

Apart from users who are usually reluctant to talk about their values, the development team may also be unwilling to share their ideas about (violated) values. Lopez et al. (P45) focus on how to encourage developers to talk about \textit{security} - especially non-technical impacts of security (e.g., how a compromise can damage trust). Lopez et al. (P45) develop a workshop format including a set of working materials to connect developers with security incidents and prompt them to speak about the consequences of such incidents. By adopting a positive, value-oriented approach toward security, the workshop first asks its attendees (developers) to read the report of a compromise (i.e., an overview of a security incident). A group of cards (e.g., value cards) are distributed among the attendees to prompt discussions about various security incidents and their impacts on stakeholders. In the next step, the attendees are encouraged to talk about the security incident from a developer's viewpoint being directly affected by the reported compromise. All this can reveal the possible impacts of the compromise on the values of the affected developer.

\subsubsection {\textbf{Evaluate and Validate Values-critical Requirements}}\label{sec:evaluateVReq}
As discussed in Section \ref{sec:CaptureValues}, many solutions have been developed to collect values from different resources. However, it is equally essential to ensure the gathered values-critical requirements (i.e., requirements that include human values) are complete and match stakeholders' needs. We found 12 primary studies (P5, P14, P16, P18, P22, P24, P26, P30, P31, P40, P42, P44) that propose solutions for this purpose. 

Lee et al. (P26) argue that there is a lack of approaches to validate the elicited requirements, particularly validating the elicited requirements against customers' inner needs (i.e., values, beliefs, and motivations that someone has but are not easily visible) and behavioral data. \textit{Customer Requirements Validation} (\textit{CuRV}) technique (P26) leveraging the mental model technique can be considered a new step in the requirements engineering process after Elicitation, Analysis, and Specification. The \textit{CuRV} technique first collects user behavior data, for example, from the earlier version of the software. It then creates a mental-requirements model to identify which of the elicited requirements correspond with users' mental states. Based on the mental-requirements model, it can suggest (a) there may be a need for re-eliciting requirements as some customers' behaviors or inner needs are not captured in elicited requirements. (b) It can provide insights for the requirements reviewers that there may be a need to re-analyze elicited requirements as some do not match customers' behaviors or inner needs.

The goal of \textit{Appraisal and Measurement of User Satisfaction} (\textit{AMUSE}) (P42) is to help the development team to select the best features that improve \textit{user satisfaction} for the next releases. To this end, \textit{AMUSE} first needs to measure the end-users' satisfaction with the current version of the product using a questionnaire. The responses collected from the questionnaire can reveal the extent to which the current users are satisfied with the product from five perspectives: \textit{effectiveness}, \textit{productivity}, \textit{hedonism}, \textit{trust}, and overall \textit{satisfaction}. 

The \textit{Ethics-Aware Software Engineering} framework (P44) is to attain ethical harmony in software artifacts and development process and includes five phases: \textit{articulation}, \textit{specification}, \textit{implementation}, and \textit{verification} and \textit{validation}. In the \textit{verification and validation} phase, the software is continuously monitored to ensure the software is aligned with its ethical values specifications (e.g., diversity, transparency) and detect and reveal any deviations.

In another study (P14), Colomo-Palacios et al. propose \textit{affect grid} to measure stakeholders' emotions in terms of \textit{pleasure} and \textit{arousal} on the collected requirements. 





\subsubsection {\textbf{Detect and/or Resolve Values-based Conflicts/Violations}} 
As values are subjective concepts, conflicts among values are inevitable. Moreover, values can be easily violated, but it would be hard to monitor values violations. We found 8 studies (P1, P5, P7, P9, P13, P15, P22, P41) that develop solutions for detecting and/or resolving conflicts among and/or violations in values. Table \ref{tbl:confilicts} gives an overview of these solutions. 
 \begin{table*}[ht]
    \caption{Solutions to detect/resolve values conflict/violation}
    \centering
    \footnotesize
\label{tbl:confilicts}
\begin{tabularx}{\linewidth}{l|X|X}
    \toprule
                  
        & \textbf{Conflict}
            & \textbf{Violation}           \\ 
    \toprule
\textbf{Detect}    &   \begin{itemize}
         \item Values conflicts are detected through an iterative refinement process of understanding stakeholders’ values, emotions, and motivation (\textbf{P1}).
         \item A set of questions are proposed, which answering to them helps detect values conflicts (\textbf{P5}).
         \item	A dialogical process can encourage, facilitate, and orchestrate discussions about different aspects (e.g., conflicts) of values among stakeholders (\textbf{P13}).
         \item It is suggested to use specific notations for representing values and a conflict relationship between them (\textbf{P15}).
         \item It is suggested to check the functional components of a proposed design solution to see if the proposed design solution has implemented all collected values (\textbf{P22}).
            \end{itemize}  
                    & 
            \begin{itemize}
                \item Stakeholders participating in co-design workshops are asked to provide reflections on the design and speciation of features/functions of the system under development to help reveal possible values violations (\textbf{P7}).
                \item The proposed solution leverages common reference points between policy statements and the corresponding privacy controls to recognize privacy violations (deficiencies) (\textbf{P9}).
                \item The proposed solution utilizes trust scenarios to generate trust rules in order to detect trust violations (\textbf{P41}).

            \end{itemize}
                    \\ 
    \midrule
\textbf{Resolve}  & \begin{itemize}
    \item It is suggested to clarify explicitly (the weight of) values (i.e., determining which values outweigh others) (\textbf{P22}).
\end{itemize}
&   \begin{itemize}
    \item A set of guidelines (corrective measures) is propped to address privacy violations between privacy policies and privacy controls (\textbf{P9}).
\end{itemize}      \\   \bottomrule

\end{tabularx}%
    \end{table*}

\textit{Value-based Requirements Engineering} (\textit{VBRE}) (P1) can reveal the conflicts between values using an iterative refinement process of acquiring and learning stakeholders’ values, emotions, and motivation. However, \textit{VBRE} has no recommendation and solution for solving the conflicts. The ``soft-goal'' concept in Tropos methodology can, to some extent, be used to describe values (P15). However, many important characteristics of values still are missed. For example, it may not be possible to address potential conflicts between values if presented with the ``soft-goal'' concept. Detweiler et al. (P15) argue that it is needed to have a distinct notation to present values properly, as values are different from goals. Having a notation for values makes links between values explicit, thereby detecting potential conflicts between values.

The studies (P7, P9, P41) attempt to detect possible values violations. \textit{UMLtrust} (P41) can model and monitor the relationship between participating parties and detect any violation of \textit{trust} occurring between them at the very beginning of a software development process using a set of specialized UML rules and notions (e.g., trust-use-case diagram). Anthonysamy et al. (P9) focus on detecting \textit{privacy} violation in online social networks. They propose a technique to measure the level of traceability between \textit{``privacy policies''} and \textit{``privacy controls''} in such networks and classify this relation as \textit{complete}, \textit{partial}, or \textit{broken}. 

Only two primary studies (P9, P22) go further and provide mechanisms or clues to resolve such conflicts and violations. Flanagan et al. (P22) propose a methodological approach to capture values from different sources (e.g., individuals) and integrate them into the design process. A distinct phase of this approach aims to identify and resolve values-based conflicts. It is argued that values conflicts arise when practitioners cannot implement all identified values. Flanagan et al. recommend that values conflicts can be fixed if (the weight of) values are explicitly clarified (i.e., determining which values outweigh others). The study (P9) introduces some corrective measures to address \textit{privacy} violation in online social networks.


\subsubsection{\textbf{Create Values-conscious Requirements Artifacts}} Our survey found that 14 primary studies (P3, P4, P5, P15, P18, P23, P30, P31, P36, P37, P38, P40, P41, P48, P50) develop new artifacts or extend the existing artifacts to manage (e.g., document, visualize) values at the early stage of software development. We call such artifacts values-conscious artifacts (e.g., a values-conscious model). As shown in Table \ref{tbl:artefact2}, values-conscious artifacts can be categorized into 8 groups.

\textbf{Goal Model.} We identified 8 primary studies that either use goal modeling techniques or extend them to model the concept of human values. For example, Curumsing et al. (P36) develop a set of goal-based notations to visualize and document emotional goals such as users’ feelings, expectations, and emotions. The study (P4) uses i* Goal Model to model consumer preferences for feature variations of a software product line. A consumer preference for features can stem from three factors: human needs, basic values, and consumer values. Barišić et al. (P31) extend Goal Models to capture and represent usability goals and requirements. Similarly, Giorgini et al. (P40) extend Tropos Goal Model to capture trust-based requirements.

\textbf{UML Diagram.} Uddin and Zulkernine (P41) extend UML diagrams to document \textit{trust} scenarios in the software development process. For example, use case diagram is extended and called trust-use-cases diagram, including two types of users (i.e., $<<$trustor$>>$ actor and a $<<$trustee$>>$ actor) and four forms of trust relationships (e.g., $<<$trust - service$>>$).

\textbf{User Story.} Two primary studies (P30, P37) suggest that user stories can be modified to elicit values. Romero-Chacón et al. (P30) argue the tasks related to \textit{web accessibility} in user stories should be explicitly labeled to ensure the inclusion of accessibility. However, an extra phase (task) in the Planning stage called ``identification of accessibility tasks'' needs to be added to Scrum to identify accessibility tasks. Harbers et al. (P37) argue that the current format of user stories cannot or does not have any mechanisms to elicit and capture stakeholders' values. Harbers et al. (P37) introduce the \textit{Value Story Workshop} technique to make user stories a values-conscious artifact. The \textit{Value Story Workshop} includes five steps: First, all relevant stakeholders are identified. In the next step, the values of each stakeholder group are elicited by leveraging a wide range of value-sensitive design techniques such as \textit{Envisioning Cards} \cite{friedman2012envisioning}. Next, one or more concrete situations that may positively or negatively impact each stakeholder's value are illustrated. Then, any concrete situation is mapped to a stakeholder requirement. The knowledge gained from the previous steps makes it possible to write user stories as: \textit{``As a <\textbf{stakeholder}>, I want <\textbf{stakeholder need}> in order to support <\textbf{value}>''}.
 
 \begin{table*}[ht]
    \caption{Artifacts are generated or modified for capturing and modeling values}
    \centering
\label{tbl:artefact2}
\footnotesize
\begin{tabularx}{\linewidth}{l|X|l}
    \toprule
\textbf{Artifact}                   
        & \textbf{Key Points and Included Primary Studies}
            & \textbf{\#}           \\ 
    \toprule
Goal Model    &   \begin{itemize}
         \item Values are represented as softgoals in Tropos Goal Model (\textbf{P15}). 
         \item Goal Modeling notations are extended to represent preference (‘‘nice-to-have’’) requirements (\textbf{P48}).
         \item Emotion-based Goal Model is a set of graphical notations to show emotion goals, roles, goals, and functional and non-functional requirements (\textbf{P36, P38}).
         \item Tropos Goal Model is extended to capture trust-based requirements (\textbf{P40}). 
         \item i* Goal Models are used to model consumer preferences (\textbf{P4}).
         \item Value-goal Model visualizes the connection between values and features (\textbf{P18}).
         \item Goal Models are extended to model usability goals and requirements (\textbf{P31}).
            \end{itemize}  
                    & 8        \\ 
    \midrule
UML Diagram  & \begin{itemize}
    \item Use Case diagram is extended to elicit and model security requirements  (\textbf{P50}).
    \item Four UML diagrams, including Use Case, Class, State-Machine, and Package, are extended to represent trust scenarios \textbf{(P41)}. 
\end{itemize}
&   2      \\   \hline
User Story   & \begin{itemize}
    \item User Story is extended to capture and represent values (\textbf{P30, P37}).
\end{itemize} &   2      \\   \hline
Persona   & \begin{itemize}
    \item A new section can be added to the Persona template to capture values (\textbf{P18}).
\end{itemize}   &   1      \\   \hline
Values Portrait   & \begin{itemize}
    \item It is a summary of values collected from different sources (\textbf{P3}). \end{itemize}&   1      \\  \hline                  
Value Identification Frame    &  \begin{itemize}
    \item It is a table that shows the values of each different stakeholder (\textbf{P5}). \end{itemize} &   1      \\ \hline 
Value Comparison Table     
        & \begin{itemize}
    \item It is a table that shows how each of the solutions/or existing systems (design options) reflects the captured values (\textbf{P5}). \end{itemize}
                    &   1      \\ \hline
Intended Value Statement   
        & \begin{itemize}
    \item It is a description of the identified values for a given system (\textbf{P23}). \end{itemize} & 1 \\\bottomrule
\end{tabularx}%
    \end{table*}
\textbf{Other Artifacts.} The study (P18) argues that the Persona template can be modified to document values. Two studies (P3, P18) suggest \textit{Values Portrait} and \textit{Intended Value Statement} artifacts, respectively, which both are a summary of the identified values. The \textit{Value-oriented and Culturally Informed} (\textit{VCIA}) model developed by Pereira et al. (P5) creates or adapts five artifacts to identify and manage values: \textit{``Stakeholder Identification Diagram''}, \textit{``Value Identification Frame''}, \textit{``Value Comparison Table''}, \textit{``Culturally Aware Requirements Framework''}, and \textit{eValue}. Both \textit{``Value Identification Frame''} and \textit{``Value Comparison Table''} artifacts utilize a tabular visualization to document and manage values at the early stages of software-intensive systems development.

\subsection{Design}
In total, we found 15 primary studies that develop solutions to operationalize values during the design phase. Table \ref{tbl:designsolution} provides an overview of these solutions. The `Solution' column shows the name of solutions (if any), followed by the `Phases/Activities' column indicating the phases/activities of solutions. We also show who can be involved in the design process when a given solution is applied (See the `Involved' column). Some solutions may use or be supported by other materials/techniques to accomplish their goal. We show such materials/techniques in the `Supporting Materials/Techniques' column. Finally, the possible artifacts produced or modified by a given solution are shown in the `Produced Artifacts' column. These 15 primary studies can be generally classified into the following categories.
\begin{table*}[]
\caption{An overview of solutions for operationalizing values in the design phase (STK: Stakeholder)}
\centering
\label{tbl:designsolution}
\resizebox{\textwidth}{!}{%
{\renewcommand{\arraystretch}{1.3}

        \makeatother
\begin{tabular}{l|l|l|l|l}
  \toprule
  \textbf{Solution}                                                                                   & \textbf{Phases/Activities}                                                                                                                                                                       & \textbf{\begin{tabular}[c]{@{}l@{}}Involved\end{tabular}}   & \textbf{\begin{tabular}[c]{@{}l@{}}Supporting Materials/Techniques\end{tabular}}                                                                                                                                 & \textbf{\begin{tabular}[c]{@{}l@{}}Produced/Modified Artifacts\end{tabular}}                                                                                                                                                                                                                                                                                                                                \\         \toprule
Speedplay (\textbf{P3})                                                                                      & \begin{tabular}[c]{@{}l@{}}(\textbf{1}) Prepare \\ (\textbf{2}) Co-develop \\ (\textbf{3}) Co-design\\ (\textbf{4}) Sustain\end{tabular}                                                                                            & All STK.                                                                         & \begin{tabular}[c]{@{}l@{}}(\textbf{1}) Design Thinking Techniques\\ (\textbf{2}) Action Research Methodology\\ (\textbf{3}) Agile Methodology\end{tabular}                                                                                      & \begin{tabular}[c]{@{}l@{}}(\textbf{1}) Values Portraits\\ (\textbf{2}) Core System Qualities\\ (\textbf{3}) Project Documentation\end{tabular}                                                                                                                                                                                                                                                                                  \\ \midrule
\begin{tabular}[c]{@{}l@{}} Value-oriented \\and Culturally\\ Informed (\textbf{P5})\end{tabular}
                                                                                    & \begin{tabular}[c]{@{}l@{}}(\textbf{1}) Analysis \\ (\textbf{2}) Synthesis\\ (\textbf{3}) Evaluation\end{tabular}                                                                                                          & All STK.                                                                         & \begin{tabular}[c]{@{}l@{}}(\textbf{1}) A Set of Design Guidelines\\ (\textbf{2}) A Set of Questions\end{tabular}                                                                                                                       & \begin{tabular}[c]{@{}l@{}}(\textbf{1}) Stakeholder Identiﬁcation Diagram\\ (\textbf{2}) Value Identiﬁcation Frame\\ (\textbf{3}) Value Comparison Table\\ (\textbf{4}) Culturally Aware Requirements Framework\\ (\textbf{5}) eValue\end{tabular}                                                                                                                                                                                              \\ \midrule
\textbf{P6}                                                                                                  & -                                                                                                                                                                                               & All STK.                                                                         & \begin{tabular}[c]{@{}l@{}}(\textbf{1}) Co-design Workshops\\ (\textbf{2}) Prompts\\ (\textbf{3}) Contextual Design Method\\ (\textbf{4}) Three Design Constraints\end{tabular}                                                   & (\textbf{1}) Value-Based Prompt                                                                                                                                                                                                                                                                                                                                                                                \\ \midrule
\begin{tabular}[c]{@{}l@{}}Value-Sensitive \\ Development \\ framework (\textbf{P7})\end{tabular}   & \begin{tabular}[c]{@{}l@{}}(\textbf{1}) Elicitation \\ (\textbf{2}) Development\\ (\textbf{3}) Execution\end{tabular}                                                                                                      & All STK.                                                                         & -                                                                                                                                                                                                                     & \begin{tabular}[c]{@{}l@{}}(\textbf{1}) Values View\\ (\textbf{2}) Modeling View\\ (\textbf{3}) Business View\end{tabular}                                                                                                                                                                                                                                                                                                       \\ \midrule
Value Pie (\textbf{P8})                                                                                      & -                                                                                                                                                                                               & All STK.                                                                         & \begin{tabular}[c]{@{}l@{}}(\textbf{1}) Values Structure\\ (\textbf{2}) A Set of Design Guidelines\end{tabular}                                                                                                                         & -                                                                                                                                                                                                                                                                                                                                                                                                     \\ \midrule
\begin{tabular}[c]{@{}l@{}}Value Sensitive \\ Action-Reflection \\ Model (\textbf{P12})\end{tabular}         & -                                                                                                                                                                                               & All STK.                                                                         & \begin{tabular}[c]{@{}l@{}}(\textbf{1}) Stakeholder Prompts (Values Scenarios)
\\ (\textbf{2}) Designer prompts (Envisioning Cards)
\end{tabular} & -                                                                                                                                                                                                                                                                                                                                                                                                     \\ \midrule
\begin{tabular}[c]{@{}l@{}}Values-led \\ Participatory \\ Design (\textbf{P13})\end{tabular}                 & \begin{tabular}[c]{@{}l@{}}(\textbf{1}) Emergence of Values\\ (\textbf{2}) Development of Values\\ (\textbf{3}) Grounding of Values\end{tabular}                                                                           & All STK.                                                                         & \begin{tabular}[c]{@{}l@{}}(\textbf{1}) Workshops (Inspiration Card Workshop)\\ (\textbf{2}) Discussions \\ (\textbf{3}) Dialogues\\ (\textbf{4}) Observations\\ (\textbf{5}) Interpretations\\ (\textbf{6}) Fictional Inquiry Technique \cite{dindler2007fictional}\end{tabular}             & -                                                                                                                                                                                                                                                                                                                                                                                                     \\ \midrule
\textbf{P15}                                                                                                 & -                                                                                                                                                                                               & All STK.                                                                         & (\textbf{1}) Six Design Principles                                                                                                                                                                                             & (\textbf{1})   Tropos Goal Model                                                                                                                                                                                                                                                                                                                                                                               \\ \midrule
\begin{tabular}[c]{@{}l@{}}Continual Value(s) \\ Assessment \\ Framework (\textbf{P18})\end{tabular} & \begin{tabular}[c]{@{}l@{}}(\textbf{1}) Identifying Values of Stakeholders\\ (\textbf{2}) Developing Initial Feature Model \\(\textbf{3}) Value-brainstorming Iterations\\ (\textbf{4}) Level-wise Evaluation\end{tabular} & All STK.                                                                         & (\textbf{1}) Personas                                                                                                                                                                                                          & \begin{tabular}[c]{@{}l@{}}(\textbf{1}) Extended Personas\\ (\textbf{2}) Extended Feature Models\\ (\textbf{3}) Value-goal Models\end{tabular}                                                                                                                                                                                                                                                                                   \\         \midrule
\textbf{P19}                                                                                                 & -                                                                                                                                                                                               & All STK.                                                                         & \begin{tabular}[c]{@{}l@{}}(\textbf{1}) Values Scenarios\\ (\textbf{2}) A Set of Design Guidelines\\ (\textbf{3}) A Set of Questions\end{tabular}                                                                                                & -                                                                                                                                                                                                                                                                                                                                                                                                     \\ \midrule
\begin{tabular}[c]{@{}l@{}}Value-Centred \\ Design \\ Framework (\textbf{P23})\end{tabular}                  & \begin{tabular}[c]{@{}l@{}}(\textbf{1}) Opportunity identification\\ (\textbf{2}) Design\\ (\textbf{3}) Evaluation\\ (\textbf{4}) Iteration\end{tabular}                                                                         & \begin{tabular}[c]{@{}l@{}}End-users,  \\ Developers, \\ Designers\end{tabular} & -                                                                                                                                                                                                                     & \begin{tabular}[c]{@{}l@{}}(\textbf{1}) Personas within Usage Contexts\\ (\textbf{2}) Statements of Intended Value\\ (\textbf{3}) Evaluation Criteria\\ (\textbf{4}) Evaluation Strategy\\ (\textbf{5}) Evaluation Procedures\\ (\textbf{6}) User Difficulty Reports\\ (\textbf{7}) Value Impact Assessment\\ (\textbf{8}) Value Delivery Scenarios\\ (\textbf{9}) Interaction Designs\\ (\textbf{10}) Design Products\\ (\textbf{11}) Causal Analyses\\ (\textbf{12}) Design Change Recommendations\end{tabular} \\ \midrule
\textbf{P38}                                                                                                 & \begin{tabular}[c]{@{}l@{}}(\textbf{1}) Content Analysis\\ (\textbf{2}) Model Building\\ (\textbf{3}) Model Validation\end{tabular}                                                                                        & All STK.                                                                         & \begin{tabular}[c]{@{}l@{}}(\textbf{1}) Ethnographic Data\\ (\textbf{2}) A Modified Content Analysis Process\end{tabular}                                                                                                               & (\textbf{1}) Goal Models                                                                                                                                                                                                                                                                                                                                                                                       \\ \midrule
HuValue Tool (\textbf{P39})                                                                                  & -                                                                                                                                                                                               & Designers                                                                   & \begin{tabular}[c]{@{}l@{}}(\textbf{1}) Values Model Structure \\ (\textbf{2}) 45 Value Cards\\ (\textbf{3}) 207 Picture Cards\end{tabular}                                                                                                      & -                                                                                                                                                                                                                                                                                                                                                                                                     \\ \midrule
\begin{tabular}[c]{@{}l@{}}Trust-aware \\ Software \\Development \\ Framework (\textbf{P41})\end{tabular}      & \begin{tabular}[c]{@{}l@{}}(\textbf{1}) Trust Scenarios Identification\\ (\textbf{2}) Trust Scenarios Modeling\\ (\textbf{3}) Trust Rule Implementation\\ (\textbf{4}) Trust Rules Deployment\end{tabular}                          & All STK.                                                                         & -                                                                                                                                                                                                                     & \begin{tabular}[c]{@{}l@{}}(\textbf{1}) Extended Use Case Diagram\\ (\textbf{2}) Extended Class Diagram\\ (\textbf{3}) Extended State-machine Diagram\\ (\textbf{4}) Extended Package Diagram\end{tabular}                                                                                                                                                                                                                               \\ \midrule
\textbf{P45}                                                                                                 & \begin{tabular}[c]{@{}l@{}}(\textbf{1}) Compromised Software\\ (\textbf{2}) Another Point of View\\ (\textbf{3}) Group Discussion\end{tabular}                                                                             & Developers                                                                  & \begin{tabular}[c]{@{}l@{}}(\textbf{1}) Prompting Cards\\ (\textbf{2}) Value Cards\\ (\textbf{3}) Security Incidents Reports\end{tabular}                                                                                                        &                                                                                                                                                                                                                                                                                                                                                                                                       \\ \bottomrule
\end{tabular}%
}\quad
}
\end{table*}


\subsubsection{\textbf{Integrate Values into the Design Process}}\label{sec:IntegrateVD}

We found six primary studies (P3, P5, P13, P23, P38, P41) that develop a phased value-centered design process that complements existing development or design processes to make values an integrated part of the design process.

Ferrario et al. (P3) argue that the current software development processes do not adequately meet or are unable to capture and satisfy social software engineering projects' needs. A social software engineering project aims to provide positive social changes for vulnerable communities. Ferrario et al. (P3) propose \textit{Speedplay} (\textit{Values-First Process}) to address these deficiencies. \textit{Speedplay} is a process model to capture values and leverage them as the key drivers in decision making in the software development process. \textit{Speedplay} is composed of 4 steps: \textit{prepare}, \textit{co-develop}, \textit{co-design}, and \textit{sustain}. \textit{Speedplay} integrates values as an integrated and central entity within software development through the following features: (1) using values as the key drivers for decisions; (2) working closely with all relevant stakeholders; (3) leveraging design thinking methods for visioning and problem solving; (4) using Agile methods to develop software; (5) using ``action research'' methodology to seek reflection from stakeholders; and (6) embracing uncertainties and risks by rapidly building and evaluating prototypes.


Uddin et al. (P41) develop a \textit{trust-aware software development framework} to manage trust concerns throughout the software development process. The framework includes four stages: \textit{trust scenarios identification}, \textit{trust scenarios modeling}, \textit{trust rule implementation}, and \textit{trust rules deployment}. Four UML diagrams, including use case, class, state machine, and package diagrams, are extended, and some trust rules are developed to model, implement, and monitor the identified trust scenarios.

Pereira and Baranauskas (P5) argue that both values and culture need to be considered when designing interactive systems. It is because values and culture cannot be separated. While a value indicates what is important for an individual, community, or society, culture shows the reason behind this importance. The \textit{VCIA model} (P5) includes the following design steps: \textit{analysis}, \textit{synthesis}, and \textit{evaluation}, which are supported by five artifacts and several design guidelines. The artifacts and guidelines act as supplementary materials to the existing approaches and tools. Designers leverage them to consider and address both culture and values explicitly throughout the development of interactive systems.   

Cockton (P23) introduces \textit{Value-Centred Design Framework} to develop value-centered systems. The framework consists of four processes: \textit{opportunity identification}, \textit{design}, \textit{evaluation}, \textit{iteration}. In each process, the development team (e.g., developers, designers) needs to perform a set of activities, which result in constructing artifacts. For example, in the \textit{design} process, the designer performs the ``value delivery scenario authoring'' activity to transform the ``values statements'' created in the \textit{opportunity identification} process to the ``value delivery scenarios'' artifact. The ``value delivery scenarios'' present how the proposed design satisfies the values captured in ``values statements''. In the \textit{evaluation} process, the designer appraises the extent to which the values delivered in the design may expose difficulties for the user, leading to the ``value impact assessment'' artifact. The activities conducted in and artifacts produced in the \textit{iteration} process suggest how the proposed design should be modified to remove undesirable user difficulties.

\textit{Values-led Participatory Design} (P13) as a three-phase design process aims to bring end-users', stakeholders', and designers' values into the design process. In each phase, the process provides several scaffolds. The first phase leverages different techniques such as workshops and presentations to foster the emergence of values with participants. In the second phase, a dialogical process is employed to encourage participants to think and discuss how the emergent values can be implemented in a new way in the design of the new software products (i.e., re-conceptualizing values). The last phase (\textit{grounding of values}) ensures that the re-conceptualized values are embedded in the final design.




\subsubsection{\textbf{Increase the Awareness of the Designer about Values}}
The decisions made by the designer to shape a software system may have implications on the values of individuals, organizations, and societies \cite{aldewereld2015design}. However, it is a challenge for the designer to recognize and think about values when designing software because values are high-level abstract concepts. We found 10 primary studies (P6, P7, P8, P12, P15, P18, P19, P39, P41, P45) that develop design materials, scaffolds, or design principles to increase awareness about values and equip designers when dealing with values during the design process.

Both \textit{Value-Centred Design Framework} (P23) and \textit{Values-led Participatory Design} (P13) discussed in Section \ref{sec:IntegrateVD} do not provide a value list to help the designer identifying values from different resources. In contrast, Kheirandish et al. (P39) develop a comprehensive values model structure supported by \textit{HuValue} tool to increase the awareness of the designer about human values. \textit{HuValue} accomplishes this goal by equipping the designer with a set of easily understandable design materials, including a values model structure with nine value clusters, 45 value cards, and 207 picture cards. These design materials help the designer bring and include the extracted values in any conversations during the design process.


It is difficult for practitioners to anticipate the long-term and systemic influences of interactive systems on people, society, and the natural environment (P19). Further to this, the current interaction design methods have no mechanisms to support designers in this regard. Nathan et al. (P19) recommend four envisioning criteria (\textit{stakeholder}, \textit{time}, \textit{value}, and \textit{pervasiveness}) to help designers recognise, reflect, and examine the long-lived consequences of interactive systems on individuals and society. The \textit{stakeholder} criterion focuses on identifying the impacts of interactive systems on directly and indirectly affected stakeholders. The \textit{time} criterion informs the designer about the longer-term implications of their work. The designer can figure out the positive and negative (if any) of their work through the \textit{value} criterion. The \textit{pervasiveness} criterion increases the awareness of the designer of the environment that interactive systems to be deployed. While the \textit{stakeholder} and \textit{value} criteria together support ethical and societal considerations of interactive systems, the combination of \textit{time} and \textit{pervasiveness} criteria presents a future-oriented viewpoint.


\textit{Value Sensitive Action-Reflection Model} (P12) leverages the ideas of co-design spaces and reflection-on-action to bring values to the technology-centric co-design process. First, the model prompts the participants of a co-design process to generate initial ideas (designs) for a design problem. Second, it uses two types of prompts to encourage the participants to elaborate on, evaluate, and reflect on their initial designs and apply the required changes (if any) to their initial designs. The first prompt, \textit{stakeholder prompt}, employs values scenarios to draw the participants' attention to the special socio-technical context of yet-to-be-built tools use (generating new features that are aligned with human values). The second one, \textit{designer} \textit{prompt}, uses \textit{Envisioning Cards} to draw attention to the more general social and contextual concerns that are readily neglected. The prompts increase (1) the number of design ideas/solutions and (2) lead to divergent thinking. The new ideas created/influenced by two types of prompts are expected to consider users' values of yet-to-be-built tools' users.

Effectively addressing human values is a good indicator of the acceptance of socio-technical systems. This type of system is necessary to achieve \textit{societal sustainability}. The current software engineering methodologies do not have any mechanisms to integrate and consider values in such systems. Barn et al. (P6) propose the \textit{co-design workshop} technique to extract value-sensitive concerns and requirements and then incorporate the extracted value-sensitive concerns into the design. The \textit{co-design workshop} technique emphasizes that all key actors, such as designers and (indirect and direct) stakeholders, should engage in building or evolving design features of the system under development. The design of a feature may lead to discussions that reveal value concerns (e.g., breaching privacy) or address values concerns. The \textit{co-design workshop} technique captures such discussions using the concept of \textit{value-based prompt}. The \textit{co-design workshop} technique utilizes contextual design method to solve a \textit{value-based prompt}.

\subsection{Implementation}






\subsubsection{\textbf{Generate Values-conscious Codes}}
Our analysis shows that five studies (P27, P29, P32, P49, P51) develop approaches that enable developers to create code or software components aligned with some human values (such code is referred to as ``values-conscious code''). 

Two studies (P27, P32) propose a domain-specific language, followed by a runtime-checking technique, with which developers can define and verify some values-related rules and specifications when they are coding. The goal is to guarantee that the codes or user interfaces developed by developers do not violate or neglect human values. Albarghouthi et al. (P32) develop a specification language to define customized fairness specifications in the code for sensitive decision-making procedures (functions). A runtime-checking technique, similar to assertions in traditional testing, is applied to check if the decisions made by a fairness-sensitive procedure violate defined fairness specifications in the code. The \textit{Rule-Based Generation of Mobile User Interface} (\textit{RUMO}) framework (P27) includes a domain-specific language that enables software engineers to define and create a set of rules and constraints to generate similar user interfaces for different platforms (i.e., \textit{usability}). A rule engine checks if the user interfaces produced for different platforms meet the defined rules and constraints (e.g., usability-related rules and constraints).

Compared to Albarghouthi et al. (P32), who introduce the \textit{fairness-aware programming} technique to consider \textit{fairness} as a first-class concern in the code, Mougouei (P51) does not focus on any specific types of values and develops the \textit{AIR} framework to support the concept of `value programming'. The \textit{AIR} framework consists of four components: \textit{``value annotation of APIs}'', \textit{``value annotation of code}'', \textit{``value inspection}'', and \textit{``value recommendation}''. The first two components identify and annotate which APIs and parts of the code are relevant to human values. The \textit{``value inspection}'' component aims to detect values breaches and violations in the code. Finally, the \textit{``value recommendation}'' component provides recommendations to alleviate or fix the detected values breaches and violations. Although the AIR framework has not been evaluated yet, it is expected to help developers learn about user values and write codes aligned with user values.



Similar to the study (P27), Rathnayake et al. (P29) try to embed human values into user interfaces. They specifically target \textit{adaptivity} and \textit{usability} as two human values. They introduce a development framework to generate an adaptive user interface automatically. This is achieved by a deep analysis of user behavior patterns and customizing web user interfaces, which are supported by machine learning solutions. Capturing user behavior is done by considering the user's mouse point waiting time and click count in each component on a webpage. The development framework detects which components/subcomponents of a web page are rarely used and dynamically switches them off based on the collected user behavior data. This improves the \textit{usability} of the website as non-technical users with minimum configuration effort can perform that.

\subsection{Testing}

\subsubsection{\textbf{Detect and Test Values-critical Features}}


This category includes solutions aiming to inspect if the values elicited from stakeholders or other resources (e.g., requirements documents) are being reflected in a designed prototype or implemented system (P5, P17, P18, P22, P42, P46, P49, P51). The \textit{VCIA} model (P5) introduces the \textit{eValue} artifact and practical guidelines on how to use this artifact, which help practitioners (e.g., designers) evaluate and reason about a software solution and its features from the perspective of values supported or ignored. The \textit{eValue} artifact helps practitioners explicitly assess the impact of the neglected value(s) and provides suggestions to identify and implement new features to include the neglected values in the next release.

Galhotra et al. (P46) develop a \textit{fairness} testing approach called \textit{Themis} to automatically measure two types of \textit{discriminations} in a software system that makes decisions (e.g., loan software): group discrimination and causal discrimination. Group discrimination focuses on detecting and scoring the difference between two or more input groups resulting in a similar output. Causal discrimination checks if the software remains fair when some characteristics (e.g., the race of individuals in the loan system) of inputs are changed.

Tramer et al. (P49) propose the \textit{FairTest} tool to aid software practitioners to identify and fix ``fairness bugs''.  The \textit{FairTest} tool takes several user attributes, including protected attributes (e.g., race), and the outputs produced by a given software system for users as input and generates an association bug report. A typical association bug report includes statistically significant associations between protected attributes and outputs, and developers can easily understand and interpret it to determine real bugs that require fixing from reported associations.

As discussed in Section \ref{sec:evaluateVReq}, \textit{AMUSE} (P42) can reveal the level of \textit{user satisfaction} with the features of the current version of a product. \textit{AMUSE} further specifies if the users perceive the product weak from the following quality aspects: \textit{effectiveness}, \textit{productivity}, \textit{hedonism}, and \textit{trust}. \textit{AMUSE} evaluates the new features that are supposed to be included in the next release (i.e., Feature Appraisal). The evaluation is done by measuring how well each candidate feature improves the \textit{effectiveness}, \textit{productivity}, \textit{hedonism}, \textit{trust}, and overall \textit{satisfaction} of the new product. \textit{AMUSE} helps practitioners decide which features to be included in the next releases for improving user satisfaction (called Feature Prioritization).
The \textit{Continual Value(s) Assessment (CVA)} framework (P18) is composed of four components: \textit{``identifying values of stakeholders}'', \textit{``developing initial feature model}'', \textit{``value-brainstorming iterations}'', and \textit{``level-wise evaluation}''. The CVA helps practitioners build the rationale from highly abstract concepts of human values to design choices of a system that the practitioners develop. CVA uses a combination of Goal Modeling and Feature Modeling techniques to continuously evaluate values at different development levels such as conceptual, behavioral, and implementation levels. This helps decision-makers pick the values they want to implement and ensure that the values are being implemented across all development levels.

\subsection{Team Organization}

\subsubsection {\textbf{Measure Team Members Values}}
Any organization may have internal and unique values expected to be accepted, respected, and followed by its staff and job applicants. However, software organizations usually do not seek to what extent their personnel and job applicants understand, agree with, and respect universal human values (e.g., inclusiveness, diversity). Hussain et al. \cite{9261980} refer to \textit{``hiring staff with values in their minds}'' (i.e., values-mined staff) as a cultural approach to address values in software. Understanding how developers think about and appreciate values can reveal any conflicts between a project's values and the project's assigned staff's values. This also may show what areas of improvement need to be sought for an organization's staff. We found five primary studies (P2, P25, P28, P33, P43) that develop solutions to understand and measure team members' perceptions of human values.

As discussed earlier, values are a subjective concept. Winter et al. (P2) argue that this implies that software engineers have different viewpoints on each value. Hence their decisions throughout a project's lifecycle are hugely influenced by their perspectives on values. Winter et al. (P2) develop \textit{V-QS} to uncover a set of manageable and understandable values patterns from diverse software engineers' values. To do this, \textit{V-QS} asks software engineers to sort a set of value statements customized for software engineering onto a grid according to what extent they agree with or like each statement. The value statements are developed by considering the ACM Code of Ethics and the 19 Schwartz value types. 

Amreen et al. (P28) assert that existing Expertise Browser tools (tools that identify experts with desired expertise) do not consider how impactful or important a developer is in the open-source community. This may lead to a fake \textit{reputation} for developers or developers do not \textit{trust} each others' profiles. \textit{Developer Reputation Estimator} (\textit{DRE}) (P30) addresses this issue by building a trustable profile for developers. \textit{DRE} collects and considers both quantity and quality measures for this purpose: (1) It calculates the number of commits of a developer; (2) It assesses the impact of the works done by a developer (e.g., it calculates how many times other developers reuse code written by developers); and (3) It measures the importance and impact of a developer's collaborators.

Ying et al. (P33) propose a reviewer recommendation technique (\textit{EARec}) to help core developers decide who should review an incoming Pull Request (PR). \textit{EARec} simultaneously considers developer (reviewer) expertise and \textit{authority} when making the decision. The level of expertise of a reviewer for each incoming PR is assessed by determining the similarity between the title and description of an incoming PR and the title and description of the PRs commented on by the reviewer. \textit{EARec} measures authority by calculating the number of PRs that reviewers have commented on together. In the last step, Random Walk with Restart (RWR) algorithm is applied to balance expertise and authority. A reviewer with many relationships with others is considered as an experienced reviewer and with more authority.

Values are not usually taken into account when organizing a software development team. A compatible team can better pursue and realize the team's goals (e.g., developing software systems aligned with today’s diverse society).
Surian et al. (P25) propose \textit{Developer-Project-Property} (\textit{DPP}) to find a list of compatible developers based on historical data. To this end, the \textit{DPP} approach first collects data regarding developers and the projects that they worked on. For each project, the approach collects two properties: (1) the programming languages used in the project and (2) the category of the project. Then a graph will be built with three nodes: Developer, Project, and Project Properties. In the last step, the RWR algorithm is applied to compute the similarity between developers.

\subsubsection{\textbf{Adjust existing Roles and Responsibilities}}
\begin{table*}[]
\caption{Roles and responsibilities for operationalizing values in software}
\centering
\label{tbl:rolesresps}
\footnotesize
\resizebox{\textwidth}{!}{%
{\renewcommand{\arraystretch}{1.1}
\begin{tabular}{lll}
\toprule
\textbf{Role}                                                                     & \textbf{Responsibilities}                                                                                                                                                                                  & \textbf{Goal}                                                                                   \\ \toprule
\begin{tabular}[c]{@{}l@{}}\textbf{Software Engineer}/\\ \textbf{Developer}\end{tabular} & They should be familiar with and apply ethical analysis techniques (\textbf{P23}).                                                                                                                             & \begin{tabular}[c]{@{}l@{}}Assess the ethical \\ consequence of decisions\end{tabular} \\ \cline{2-3} 
                                                                         & \begin{tabular}[c]{@{}l@{}}They should increase their knowledge about the needs and characteristics \\ of disabled people and include them in the usability test (\textbf{P34}).\end{tabular}                & Address accessibility issues                                                           \\ \cline{2-3} 
                                                                         & \begin{tabular}[c]{@{}l@{}}They should check if all users can access and use finished functionalities \\ and fix any accessibility issues (\textbf{P30}).\end{tabular}                                           & Address accessibility issues                                                           \\ \midrule
\textbf{Product Owner}                                                            & \begin{tabular}[c]{@{}l@{}}They should prioritize accessibility from the beginning of the project and \\ produce user stories that take into account disabled people and their needs (\textbf{P34}).\end{tabular} & Address accessibility issues                                                           \\ \midrule
\textbf{Scrum Master}                                                             & They should guarantee that the DONE definition covers accessibility (\textbf{P34}).                                                                                                                              & Address accessibility issues                                                           \\ \midrule
\textbf{Validation Team}                                                          & \begin{tabular}[c]{@{}l@{}}They should use well-known criteria-based frameworks to validate and \\ audit a software product from the accessibility perspective (\textbf{P30}).\end{tabular}                       & Address accessibility issues                                                           \\ \midrule
\textbf{Co-pilot}                                                  & They ensure fairness and autonomy in software crowdsourcing platforms (\textbf{P35}).                                                                                                                            & Establish fairness and autonomy                                                      \\ \bottomrule

\end{tabular}%
}\quad
}
\end{table*}
Four studies (P21, P30, P34, P35) focus on roles and responsibilities to embed values in software. As Table \ref{tbl:rolesresps} shows, software engineers/developers, Product Owners, Scrum Masters, validation team, and co-pilots need to adjust their responsibilities to address values. Miller and Larson (P23) recommend new responsibilities and skills for software engineers to develop software services or products that are more aligned with the values of a wide range of diverse stakeholders. They emphasize that software engineers should be familiar with and perform ethical analysis techniques (e.g., utilitarian analysis and deontological analysis) to put human values at a central point in the decision-making and predict the ethical consequence of each of their decisions. 

Pellegrini et al. (P34) argue that many \textit{accessibility} issues in software projects are due to (1) postponing the implementation of accessibility features by teams that adopt Agile methods (for example, because they adopt the Minimum Viable Product approach), and (2) a lack of knowledge on the implementation of accessibility. Pellegrini et al. (P34) define a set of new responsibilities for roles involved in software development to address this issue. For example, Product Owner should prioritize accessibility from the beginning of the project and produce user stories that take into account disabled people and their needs. Scrum Master should guarantee that the DONE definition covers accessibility. 

In the same line, Romero-Chacón et al. (P30) suggest that Scrum should be adjusted to integrate \textit{accessibility} and \textit{usability} criteria in the beginning steps of developing a software project. 
The customized Scrum has three extra phases (tasks) in Sprint: \textit{``accessibility test}'', \textit{``accessibility fixes}'', and \textit{``accessibility review}''. During the \textit{``accessibility test}'' phase, developers check if all users can access and use finished functionalities. Any accessibility issues need to be resolved in the \textit{``accessibility fixes}'' phase by developers. Once accessibility corrections are applied, the validation team, with the help of customers, determines which web pages and to what extent should be evaluated from the accessibility perspective. Next, the validation team uses the Web Content Accessibility Guidelines (WCAG) as a criteria-based framework to audit the entire website (or a representative sample of the website).





\section{A Further Analysis of Current Solutions}\label{sec:solutionsanalysis}
In Section \ref{sec:taxonomy}, we grouped the 51 solutions into ten categories based on five areas in software engineering: requirements, design, implementation, testing, and team organization. This section further classifies and articulates the 51 identified solutions from two perspectives: the type of human values they aim to operationalize and tool support.

\subsection{Operationalized Human Values}\label{sec:targtedvalues}
Our analysis shows that the 51 solutions can be arranged into two groups according to the type of human values operationalized by them: Holistic View and Exclusive View.
\subsubsection{Holistic View}
\begin{figure*}
   \centering
    \includegraphics[width=0.65\textwidth]{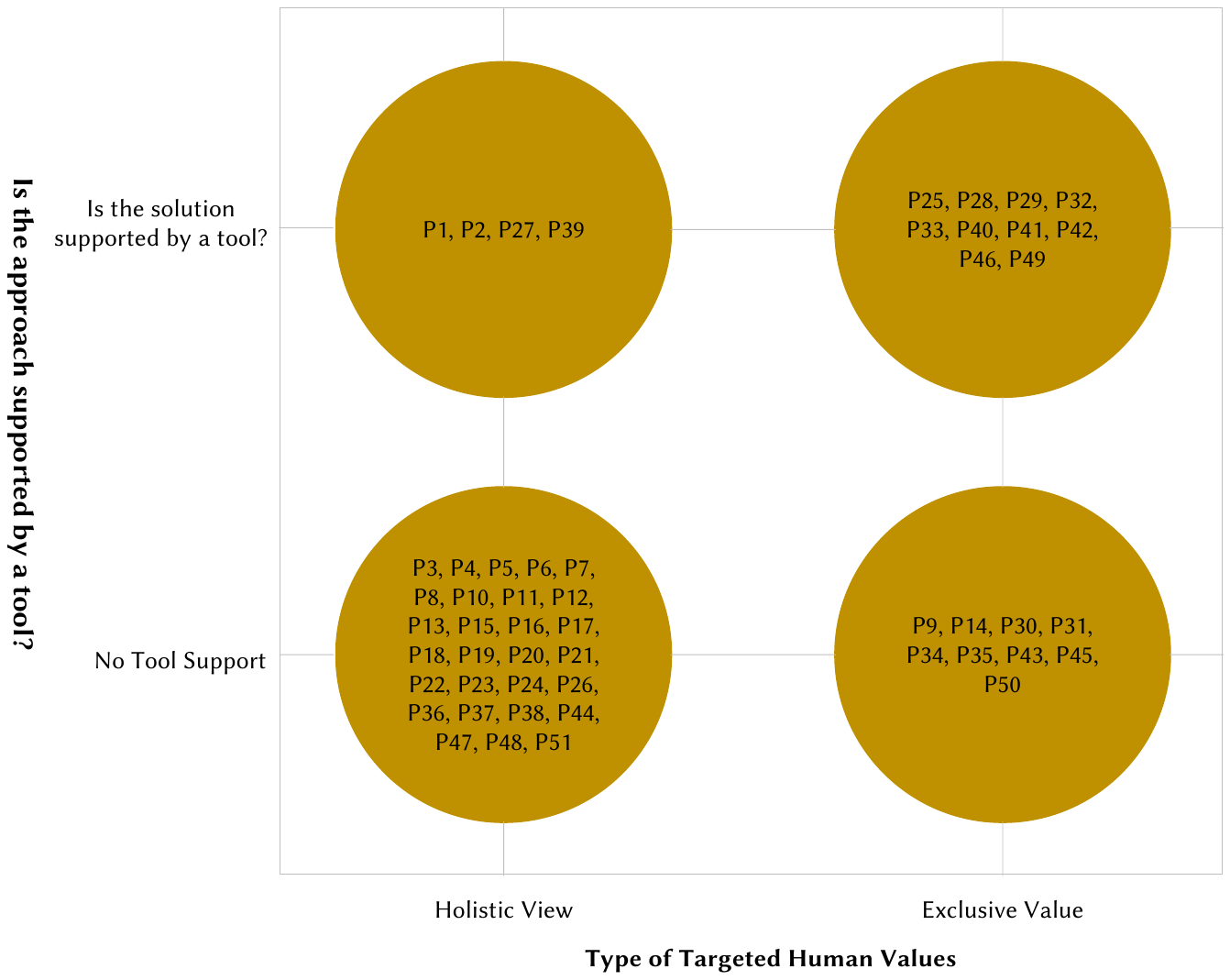}
    \caption{Distribution of primary studies per the type of operationalized values and tool support.}
    \label{fig:targetvaluestools}
\end{figure*}

We found that the majority of the primary studies (32 out of 51, 62.7\%) do not focus on specific values and provide solutions that aim to operationalize human values as a concept that refers to a wide range of human-centric issues (e.g., emotional value and economic value) without explicitly distinguishing them. In Table \ref{tbl:AppToolsValues} and Figure \ref{fig:targetvaluestools}, we label such solutions as a `Holistic View' (See `Values' column in Table \ref{tbl:AppToolsValues}). For example, while P12 focuses on safety, youth, and addressing homelessness to show the effectiveness of the \textit{Value Sensitive Action-Reflection} model, the model is not restricted to any specific values. It can capture and reveal any human values during the design process and recognize (new) features that are more aligned with human values. Similarly, the \textit{RUMBO} (\textit{Rule-Based Generation of Mobile User Interface}) framework (P27) can be used to define any human values and detect any value violations, but P27 mostly targets accessibility and usability to evaluate the \textit{RUMBO} framework. 


Some studies (e.g., P2, P4, P16) use the well-known values models (e.g., the Schwartz theory of basic values) as a starting point to show and assess the functionalities of their proposed solutions. 
We also categorized such solutions (e.g., \textit{Values Q-sort} (P2)) as `Holistic View'. 
\subsubsection{Exclusive Values}
As shown in Figure \ref{fig:targetvaluestools}, we found 19 studies that exclusively target one or two human values. In this category, Fairness (P32, P35, P46, P49) and Trust (P28, P40, P41, P43) are targeted the most as four primary studies provide solutions to address each of these values, followed by Accessibility (P30, P34), Usability (P29, P31), and Security (P45, P50). Other targeted values are Compatibility (P25), Privacy (P9), Arousal (P14), Reputation (P28), Adaptability (P29), Pleasure (P14), Authority (P33), Satisfaction (P42), Autonomy (P35) and Productivity (P31). For example, P33 proposes a method called \textit{EARec} to consider developer authority besides expertise when assigning them to review an incoming Pull Request (PR), while P34 proposes new responsibilities for Scrum Master, Product Owner, and Team Members to implement accessibility in software projects. 
\subsection{Tool Support}\label{sec:toolsupport}
We also investigated if any tools are developed to support the proposed solutions. As shown in Table \ref{tbl:AppToolsValues} (See `Tool' column), only 14 primary studies (27.4\%) develop tools to (semi-) automate some or all parts of their respective solutions for operationalising values. Figure \ref{fig:targetvaluestools} shows that 10 of 14 of the developed tools address one or two exclusive values, and the rest works for any human values. We observe from Figure \ref{fig:targetvaluestools} that the majority of the solutions (28 out of 32) targeting any value (`Holistic View') are not supported by any tools.
\section{Implications for Research and Practice}\label{sec:discussion}




In Sections \ref{sec:demographics}, \ref{sec:taxonomy}, and \ref{sec:solutionsanalysis}, we provided an in-depth analysis of the research around operationalizing values in software and presented a taxonomy of solutions for this purpose. In the coming sections, we present implications for research and practice.
We also highlight some promising research areas and open issues that need more consideration from researchers and practitioners.




\subsection{Develop values-friendly solutions beyond requirements engineering and design}
Our findings highlight that requirements engineering and design have received a significantly higher number of contributions than software implementation and testing. Out of the identified 51 solutions, 33 relate to the requirements engineering phase, making requirements engineering the most common area to seek guidance from for operationalizing values (See Table \ref{tbl:AppToolsValues} and Figure \ref{fig:classifcation_new}). We argue that this fertility results from a combination of (1) recognition and labeling of human values as goals in social sciences and psychology dating back to the 1970's \cite{schwartz1987toward, rokeach1973nature}, and (2) the development of goal-oriented RE techniques aimed at satisfying user goals and expectations from the system more than three decades ago \cite{213081, martin1983human}. 


The concentration of solutions focused on requirements engineering and design highlighted in our results helps explain the findings by \cite{9261980} who report that the traceability link between values design and values implementation is often broken. Even the techniques from HCI (Human-Computer Interaction) that may support values consideration during the implementation phase often do not cross over to software engineering. Our taxonomy of solutions in Figure 7 clearly highlights the need for more solutions for the implementation phase of software engineering. The current tally sits as 5 solutions for the implementation phase compared to 16 in the design phase. 


Furthermore, the proposed solutions to integrate values into the software system during requirements engineering are better supported by industrial validations than those introduced in software development or testing. 
The proposed value-based solutions for the requirements engineering phase include values elicitation techniques, guidance for evaluating and validating values-critical requirements, and approaches to develop artifacts informed by human values. The requirements engineering artifacts, such as value stories,  are of critical importance since they act as inputs to downstream activities, hence enabling the operationalization of values in software. While the availability of a healthy pool of techniques and guidelines during the RE phase is encouraging, further contributions are still needed for the development and testing phases to ensure the successful implementation of values into the software (as discussed in Section \ref{sec:solutionsneed}).

\subsection{Collaborate with practitioners to co-develop and co-validate values-based tools}


Practitioners mainly rely on tools to build technology. When it comes to sensitive issues like values, the developers need tools supporting the development and implementation of the target values. Still, more importantly, their effectiveness needs to be pre-validated, ideally in a commercial project or setting. Based on our examination of the literature, we highlight this as an open area, requiring more attention and effort. Potential growth in the availability of tools can profoundly increase the incorporation of the specified values, as we have seen in the case of privacy, security, and accessibility that enjoy mature research and implementation tools support  \cite{vigo2013benchmarking,story2021awareness}. Our examination of the trend in tools development suggests that most of the proposed techniques in our sample of studies are either not supported by a tool or the accompanied tools are not validated in real-life projects. This creates a multi-faceted problem. a) It takes years before such innovations are commercially feasible and reach the industry (this is a general trend in software engineering research). b) It would be challenging to prove their efficacy and efficiency in real-life projects. c) Since these tools are perceived as `cooked-up' in research labs rather than co-developed with the practitioners, tools uptake requires a disruption rather than diffusion of the innovation \cite{rogers2010diffusion}.
Collaborative approaches such as co-design suggested by the research on human values in software engineering can be leveraged to improve the development and acceptance of tools designed to integrate values in software (discussed further in Section \ref{sec:solutionsneed}).
\subsection{Align Developer values with the purpose of technology development}


Technologies actively help shape our societies, rather than being neutral means to help realize human ends \cite{verbeek2008morality}. Studies in software engineering \cite{9261980}, machine learning \cite{crawford2016artificial}, computer game design \cite{flanagan2014values} and social sciences \cite{winner1980artifacts} have adequately demonstrated that technologies can (by design or accident) embody values, especially of those who create them. Therefore, the role and power of the creators of these technologies at the individual level and team level cannot be overemphasized.

Our findings, however, highlight that only a handful of solutions exist to identify or measure developer team values. Close examination of these studies also reveals that no attempt has been made to link or align the identification of team and developer values with the technology's development purpose or intentions. Similar to other studies in software engineering \cite{winter2019advancing, 9261980}, our findings suggest that building values awareness in the team members, creating values-specific roles for team members, and aligning developer motivations with values expectations from technology is likely to result in achieving the desired software outcomes. We believe that the `friction' caused by the misalignment of developer values with system intentions can severely hamper building technologies with the desired purpose and impact. Given the importance of the topic and paucity of research to identify, measure and align developer values with the purpose of technology calls for a significant investment of research efforts in this area.

 
\subsection{Develop metrics for values validation and measurement}\label{sec:solutionsneed}


Metrics are a vital part of the verification and testing of software design and code artifacts \cite{hailpern2002software,farooq2011software}. Although it is argued that values operationalization requires the development of value-specific metrics to support values-based testing and verification  \cite{mougouei2018operationalizing}, our analysis reveals that there has been little progress on this front. While our survey identified a few mechanisms that measure team member values, the metrics that measure individual values are virtually non-existent. A lack of consensus on the definitions of values could be offered as a possible explanation for this phenomenon. Even if this justification is entertained, the implementation of values-critical requirements or their quality is particularly difficult to verify without the availability of metrics.  Given the subjective nature of values, a general lack of availability of quantitative metrics is hardly surprising; however, not finding qualitative measures presents an opportunity for contribution and clearly suggests that the software engineering research community needs more focused efforts to address this problem.

\section{Limitations and Threats to Validity}\label{limiations}


The paper collection and data analysis steps may have introduced some limitations and threats to our survey. Following the recommendations proposed in \cite{jalali2012systematic, webster2002analyzing}, we first created an initial pool of papers by executing a very simple but broad keyword-based search string on Google Scholar. We then performed the backward snowballing technique to minimize missing the relevant papers. As we discussed in Section \ref{sec:papercolection}, we did not conduct the forward snowballing technique. Hence, we acknowledge that our final set of primary studies may not be the most comprehensive one, and we might have missed some important primary studies. Some steps, such as reducing 315 papers to 131 papers and the identification of papers for the snowballing iteration, in the paper collection process (See Section \ref{sec:papercolection}) were performed by the first author. Such activities may have introduced a subjectivity threat. Our strategy to minimize such a threat was to maintain a shared Excel spreadsheet file and record all the reasons for including and excluding papers in each step of the paper collection process. This made the paper collection process for all authors visible and enabled them to review the chosen and excluded papers and provide their feedback.

We have classified the 51 primary studies for different purposes (Sections \ref{sec:demographics}, \ref{sec:taxonomy}, and \ref{sec:solutionsanalysis}). To minimize the misclassification of the primary studies in Sections \ref{sec:researchtype} and \ref{sec:targtedvalues}, we tried to reduce our personal judgments and interpretations when analyzing the collected data. Concerning the taxonomy of solutions (Section \ref{sec:taxonomy}), the first author manually analyzed the collected qualitative data (data items D14 to D17) for this purpose. In order to reduce the possible subjective bias in building the taxonomy, the taxonomy was built in several iterations and constantly shared with other authors to seek their feedback. 
\section{Conclusion and Future Work}\label{sec:conclusion}
This survey has provided a detailed analysis of the state-of-the-art solutions for operationalizing values in software. Our findings will allow software engineering practitioners and researchers to understand the research around operationalizing values in software and provide some essential open research areas for research and investments. The main results of our survey are:
\begin{itemize}

    \item We provided a taxonomy of 51 solutions for operationalizing values in software. At the highest level, the taxonomy groups the 51 solutions into five software engineering areas: requirements, design, implementation, testing, and team organization. The 51 solutions are further classified into ten `not mutually exclusive' groups.
    \item We observed that most of the solutions attempt to help operationalize values in the areas of requirements (33 solutions) and design (15 solutions). Consequently, our survey only found a few solutions that contribute to other areas in software engineering: implementation (5 solutions), testing (8 solutions), and team organization (9 solutions). 
    \item Concerning the type of operationalized values, the 51 solutions are classified into two categories: holistic view and exclusive values. While the former category includes solutions (32, 62.7\%) that attempt to embed any types of human values in software, the solutions (19, 37.2\%) in the latter category exclusively target one or two human values (e.g., fairness).
    \item We found a lack of tool support for the identified solutions, as only 14 out of the 51 solutions (27.4\%) are supported by tools.
    
\end{itemize}

In this study, we focused on solutions for operationalizing human values in software engineering. There are several future directions for this work. It is worth identifying and analyzing techniques proposed in the AI community that attempt to operationalize human values in machine learning algorithms and models. Given the increasing importance of values consideration in software among software practitioners, multi-vocal literature reviews \cite{garousi2016and} can be conducted on grey literature (e.g., practitioners' blogs, white papers) to identify further solutions to incorporate values in software. Finally, future review studies can investigate how values are addressed in non-software systems. Such studies can provide insights into how human values are operationalized outside the software realm and maybe transfer some of these insights into software engineering.

\appendices
\section*{Acknowledgment} This work is supported by ARC Laureate Fellowship FL190100035.
\section{An Overview of 51 Solutions}\label{sec:appendix}
Table \ref{tbl:AppToolsValues} provides an overview of all 51 solutions. The `Method Description' column indicates the name of each solution (if any) and provides a summary of them. The `Tool' column indicates if solutions are accommodated by any tool. The `Values' column shows if solutions attempt to operationalize specific human values or target any human values. The `Category' column shows each solution belongs to which of the 10 categories (\textbf{O1} to \textbf{O10} in Figure \ref{fig:classifcation_new}).


\begin{table*}[]
\caption{Solutions, Tooling Support, Operationalized Values, and Categories}
\centering
\label{tbl:AppToolsValues}
\resizebox{\textwidth}{!}{%
{\renewcommand{\arraystretch}{1.14}
\begin{tabular}{|c|l|c|c|c|}
\hline
\textbf{ID} & \textbf{Solution Name and Description}                                                                                                                                                                                                                                                  & \textbf{Tool} & \textbf{Values}                                                    & \textbf{Category}                                                        \\ \hline
P1          & \textbf{Value-based Requirements Engineering (VBRE)} supports identifying stakeholders' values and motivations.                                                                                                                                                                                  &       \checkmark        & Holistic View                                                      & (O1) (O3)                                                                \\ \hline
P2          & \textbf{Values Q-sort (V-QS)} enables the identification and explanation of values at the system, personal, instantiation levels.                                                                                                                                                                &       \checkmark        & Holistic View                                                      & (O9)                                                                     \\ \hline
P3          & \textbf{Speedplay (Values-First Process Model)} integrates principles of values research to software engineering practices.                                                                                                                                                                      & -             & Holistic View                                                      & (O1) (O4) (O5)                                                           \\ \hline
P4          & \textbf{Consumer Preference Meta-Model (CPMM)} utilizes the preferences (values) of end-users to capture requirements.                                                                                                                                                                           & -             & Holistic View                                                      & (O1) (O4)                                                                \\ \hline
P5          & \textbf{Value-oriented and Culturally Informed (VCIA)} supports the consideration of values and culture in interactive systems design.                                                                                                                                                           & -             & Holistic View                                                      & \begin{tabular}[c]{@{}c@{}}(O1) (O2) (O3) \\ (O4) (O5) (O8)\end{tabular} \\ \hline
P6          & \textbf{It (unnamed)} translates value sensitive concerns to software engineering processes/practices.                                                                                                                                                                                           & -             & Holistic View                                                      & (O1) (O6)                                                                \\ \hline
P7          & \begin{tabular}[c]{@{}l@{}}\textbf{Value-Sensitive Software Development (VSSD) Framework} specifies and analyses a software system from a value-sensitive \\ viewpoint by linking the values, the domain characteristics, and the resulting system.\end{tabular}                                 & -             & Holistic View                                                      & (O3) (O6)                                                                \\ \hline
P8          & \textbf{Value Pie (VP)} enables designers to have a comprehensive and informed understanding of values during the design.                                                                                                                                                                        & -             & Holistic View                                                      & (O6)                                                                     \\ \hline
P9          & \textbf{It (unnamed)} measures the level of traceability between ``privacy policies'' and ``privacy controls'' in social networks.                                                                                                                                                                & -             & Privacy                                                            & (O3)                                                                     \\ \hline
P10         & \textbf{It (unnamed)} predicts the values of a group of users working in an environment by soliciting their task preferences.                                                                                                                                                                  & -             & Holistic View                                                      & (O1)                                                                     \\ \hline
P11         & \begin{tabular}[c]{@{}l@{}}\textbf{It (unnamed)} elicits the human values of stakeholders when working in their environments and visualizes the identified\\ information using a visual mapping tool to effectively communicate human values.\end{tabular}        & -             & Holistic View                                                      & (O1)                                                                     \\ \hline
P12         & \textbf{Value Sensitive Action-Reflection Model }supports different types of value-sensitive envisionings during the design.                                                                                                                                                                     & -             & Holistic View                                                      & (O1) (O6)                                                                \\ \hline
P13         & \begin{tabular}[c]{@{}l@{}}\textbf{Values-Led Participatory Design (Dialogical Process)} cultivates the emergence of values using dialogues, supports its \\ development, and grounds them into current design practice.\end{tabular}                                                            & -             & Holistic View                                                      & (O1) (O3) (O5)                                                           \\ \hline
P14         & \textbf{Affect Grid} characterizes and measures emotional requirements in the software development process.                                                                                                                                                                                      & -             & \begin{tabular}[c]{@{}c@{}}Arousal, \\ Pleasure\end{tabular}       & (O2)                                                                     \\ \hline
P15         & \textbf{Unnamed}. Some additions (e.g., notations) are suggested to be added to Tropos methodology to capture and represent values.                                                                                                                                                              & -             & Holistic View                                                      & \begin{tabular}[c]{@{}c@{}}(O1) (O3) (O4) \\ (O6)\end{tabular}           \\ \hline
P16         & \textbf{It (unnamed)} elicits values and specifies their relationships with requirements with the users' evaluation of features.                                                                                                                                                              & -             & Holistic View                                                      & (O1) (O2)                                                                \\ \hline
P17         & \begin{tabular}[c]{@{}l@{}}\textbf{Values-based Software Testing} enables the tester to determine which features should be tested first based on three characteristics \\ of features obtained from the clients and the market: business importance, quality risk, and cost.\end{tabular}        & -             & Holistic View                                                      & (O8)                                                                     \\ \hline
P18         & \begin{tabular}[c]{@{}l@{}}\textbf{Continual Value(s) Assessment (CVA) Framework} guides the software team to continuously consider and evaluate values at \\different development levels such as conceptual, behavioral, and implementation levels. \end{tabular}                                                                                                                                                      & -             & Holistic View                                                      & \begin{tabular}[c]{@{}c@{}}(O1) (O2) (O4) \\ (O6) (O8)\end{tabular}      \\ \hline
P19         & \begin{tabular}[c]{@{}l@{}}\textbf{It (unnamed)} recommends four envisioning criteria (stakeholder, time, value, and pervasiveness) to help designers \\recognise, reflect, and examine the long-lived consequences of interactive systems on individuals and society.\end{tabular}             & -             & Holistic View                                                      & (O6)                                                                     \\ \hline
P20         & \begin{tabular}[c]{@{}l@{}}\textbf{It (unnamed)} introduces and confirms some guidelines, which complements the RE process, to help \\the requirements engineer understand and recognise the emotional issues and values of a computer system's end-users.\end{tabular}                         & -             & Holistic View                                                      & (O1)                                                                     \\ \hline
P21         & \textbf{It (unnamed)} suggests that software engineers should learn and apply ethical analysis techniques (e.g., utilitarian analysis).                                                                                                                                                          & -             & Holistic View                                                      & (O10)                                                                    \\ \hline
P22         & \textbf{It (unnamed)} enables designers to discover, analyse, and integrate values during the design process.                                                                                                                                                                                    & -             & Holistic View                                                      & \begin{tabular}[c]{@{}c@{}}(O1) (O2) (O3) \\ (O8)\end{tabular}           \\ \hline
P23         & \begin{tabular}[c]{@{}l@{}}\textbf{Value-Centred Design Framework} supports value-centred design by structuring system development around four \\processes consisting of artifacts (e.g., value delivery scenarios) and activities (e.g., value operationalization).\end{tabular}               & -             & Holistic View                                                      & (O1) (O4) (O5)                                                           \\ \hline
P24         & \begin{tabular}[c]{@{}l@{}}\textbf{Unnamed}. A values-focused step, Human Value Analysis, to be added to domain modeling guidelines to help the analyst \\learn the real expects of users from the system to avoid the possible system rejection and negative social implications.\end{tabular} & -             & Holistic View                                                      & (O2)                                                                     \\ \hline
P25         & \begin{tabular}[c]{@{}l@{}}\textbf{Developer-Project-Property (DPP) Graph} finds a list of compatible developers based on historical data regarding \\developers and the projects that they worked on.\end{tabular}                                                                             &    \checkmark           & Compatibility                                                      & (O9)                                                                     \\ \hline
P26         & \textbf{Customer Requirements Validation (CuRV)} elicits users' inner needs through the mental model.                                                                                                                                                                                            & -             & Holistic View                                                      & (O2)                                                                     \\ \hline
P27         & \begin{tabular}[c]{@{}l@{}}\textbf{Rule-Based Generation of Mobile User Interface (RUMO)} enables building similar user interfaces for different \\platforms (e.g., iOS, Android) to help users feel the same user experience on different platforms.\end{tabular}                              &               & Holistic View                                                      & (O7)                                                                     \\ \hline
P28         & \begin{tabular}[c]{@{}l@{}}\textbf{Developer Reputation Estimator (DRE)} finds collaborators for OSS projects by considering the impact of \\contributors' works besides their technical networks.\end{tabular}                                                                                 &    \checkmark           & \begin{tabular}[c]{@{}c@{}}Reputation, \\ Trust\end{tabular}       & (O9)                                                                     \\ \hline
P29         & \begin{tabular}[c]{@{}l@{}}\textbf{It (unnamed)} supports automatically generating an adaptive user interface.\end{tabular}                                                                   &      \checkmark         & \begin{tabular}[c]{@{}c@{}}Adaptivity, \\ Usability\end{tabular}   & (O7)                                                                     \\ \hline
P30         & \textbf{Customised Scrum Methodology} supports building accessible software products in a more agile and flexible fashion.                                                                                                                                                                       & -             & Accessibility                                                      & (O2) (O4) (O10)                                                          \\ \hline
P31         & \textbf{It (unnamed)} supports Domain-Specific Language development with a focus on usability concerns.                                                                                                                                                                                          & -             & \begin{tabular}[c]{@{}c@{}}Productivity, \\ Usability\end{tabular} & (O2) (O4)                                                                \\ \hline
P32         & \textbf{Fairness-aware Programming} declaratively specifies fairness definitions in the decision-making code and checks them at runtime.                                                                                                                                                         &      \checkmark         & Fairness                                                           & (O7)                                                                     \\ \hline
P33         & \textbf{EARec} simultaneously considers reviewer expertise and authority when assigning a reviewer to an incoming Pull Request.                                                                                                                                                                  &       \checkmark        & Authority                                                          & (O9)                                                                     \\ \hline
P34         & \textbf{It (unnamed)} defines a set of roles and responsibilities to implement accessibility in Agile projects.                                                                                                                                                                                  & -             & Accessibility                                                      & (O10)                                                                    \\ \hline
P35         & \textbf{Co-pilot Role} ensures fairness in crowdsourcing software development.                                                                                                                                                                                                                   & -             & \begin{tabular}[c]{@{}c@{}}Fairness, \\ Autonomy\end{tabular}      & (O10)                                                                    \\ \hline
P36         & \textbf{Emotion-oriented RE Technique} captures and evaluates the emotional goals of users (e.g., elderly users).                                                                                                                                                                                & -             & Holistic View                                                      & (O1) (O4)                                                                \\ \hline
P37         & \textbf{Value Story Workshop} identifies user stories that account for values.                                                                                                                                                                                                                   & -             & Holistic View                                                      & (O1) (O4)                                                                \\ \hline
P38         & \textbf{Emotional Models} help develop software-intensive systems that consider and satisfy users' emotional needs.                                                                                                                                                         & -             & Holistic View                                                      & (O1) (O4) (O5)                                                           \\ \hline
P39         & \begin{tabular}[c]{@{}l@{}}\textbf{HuValue Tool }supports designers to address human values in their designs using a printed values model structure, \\value cards, and picture cards. \end{tabular}                                                                                                                                      &       \checkmark        & Holistic View                                                      & (O6)                                                                     \\ \hline
P40         & \textbf{Secure Tropos} is a framework to capture, model, and verify (security) requirements using security-specific concepts.                                                                                                                                                                    &      \checkmark         & Trust                                                              & (O2) (O4)                                                                \\ \hline
P41         & \textbf{UMLtrust} extends UML to integrate and specify trust in the software development lifecycle.                                                                                                                                                                                              &       \checkmark        & Trust                                                              & \begin{tabular}[c]{@{}c@{}}(O3) (04) (O5) \\ (O6)\end{tabular}           \\ \hline
P42         & \begin{tabular}[c]{@{}l@{}}\textbf{Appraisal and Measurement of User Satisfaction (AMUSE)} helps requirements engineers to build user satisfaction \\into the early stage of software.  \end{tabular}                                                                                                                                       &       \checkmark        & Satisfaction                                                       & (O1) (O2) (O8)                                                           \\ \hline
P43         & \begin{tabular}[c]{@{}l@{}}\textbf{Unnamed}. Three practices in Agile methods, including ``Sprint/iteration planning'', ``Daily stand-up'', and ``Sprint/iteration''\\ retrospective, can help build trust in the software team.\end{tabular}                                                               & -             & Trust                                                              & (O9)                                                                     \\ \hline
P44         & \begin{tabular}[c]{@{}l@{}}\textbf{Ethics Aware Software Engineering Framework} is to ensure the software is aligned with its ethics specifications and detect \\and reveal any deviations in software artifacts and development processes.\end{tabular}              & -             & Holistic View                                                      & (O1) (O2)                                                                \\ \hline
P45         & \textbf{It (unnamed)} develops and uses working materials to encourage developers to talk about security implications and impacts.                                                                                                                                                               & -             & Security                                                           & (O1) (O6)                                                                \\ \hline
P46         & \textbf{Themis} detects and measures software discriminates with the focus on causality in discriminatory behaviour.                                                                                                                                                                             &     \checkmark          & Fairness                                                           & (O8)                                                                     \\ \hline
P47         & \textbf{PC-RE} is a framework to extract the characteristics of users, their personal requirements, and physical contextual properties.                                                                                                                                                          & -             & Holistic View                                                      & (O1)                                                                     \\ \hline
P48         & \textbf{It (unnamed)} extends goal modeling techniques to model, reason, and prioritise preferences requirements.                                                                                                                                                                                & -             & Holistic View                                                      & (O1) (O4)                                                                \\ \hline
P49         & \textbf{FairTest} helps developers identify and fix ``fairness bugs'' and generates an association bug report.                                                                                                                                                                               &       \checkmark        & Fairness                                                           & (O7) (O8)                                                                \\ \hline
P50         & \textbf{It (unnamed)} extends use case diagram by adding two concepts (misuse cases and misusers) to elicit security requirements.                                                                                                                                                               & -             & Security                                                           & (O1) (O4)                                                                \\ \hline
P51         & \textbf{AIR Framework} supports the concept of `value programming'.                                                                                                                                                                                                                              & -             & Holistic View                                                      & (O7) (O8)                                                                \\ \hline
\end{tabular}

}\quad
}
\end{table*}

\bibliographystyle{ieeetr}
\bibliography{ms}

\begin{thebibliography}{100}

\bibitem{whittle2021case}
J.~Whittle, M.~A. Ferrario, W.~Simm, and W.~Hussain, ``A case for human values
  in software engineering,'' {\em IEEE Software}, vol.~38, no.~1, pp.~106--113,
  2021.

\bibitem{ferrario2016values}
M.~A. Ferrario, W.~Simm, S.~Forshaw, A.~Gradinar, M.~T. Smith, and I.~Smith,
  ``Values-first se: research principles in practice,'' in {\em 2016 IEEE/ACM
  38th International Conference on Software Engineering Companion (ICSE-C)},
  pp.~553--562, IEEE, 2016.

\bibitem{sellen2009reflecting}
A.~Sellen, Y.~Rogers, R.~Harper, and T.~Rodden, ``Reflecting human values in
  the digital age,'' {\em Communications of the ACM}, vol.~52, no.~3,
  pp.~58--66, 2009.

\bibitem{kirkham2020using}
R.~Kirkham, ``Using european human rights jurisprudence for incorporating
  values into design,'' in {\em Proceedings of the 2020 ACM Designing
  Interactive Systems Conference}, pp.~115--128, 2020.

\bibitem{schwartz2012overview}
S.~H. Schwartz, ``An overview of the schwartz theory of basic values,'' {\em
  Online readings in Psychology and Culture}, vol.~2, no.~1, pp.~2307--0919,
  2012.

\bibitem{rokeach1973nature}
M.~Rokeach, ``The nature of human values.,'' pp.~359--361, 1973.

\bibitem{koch2013approximate}
S.~H. Koch, R.~Proynova, B.~Paech, and T.~Wetter, ``How to approximate users’
  values while preserving privacy: experiences with using attitudes towards
  work tasks as proxies for personal value elicitation,'' {\em Ethics and
  information technology}, vol.~15, no.~1, pp.~45--61, 2013.

\bibitem{galhotra2017fairness}
S.~Galhotra, Y.~Brun, and A.~Meliou, ``Fairness testing: testing software for
  discrimination,'' in {\em Proceedings of the 2017 11th Joint Meeting on
  Foundations of Software Engineering}, pp.~498--510, 2017.

\bibitem{incidentDB}
``{AI Incident Database}.''

\bibitem{fbCamAna}
``{Cambridge Analytica, GDPR - 1 year on - a lot of words and some action}.''

\bibitem{neate_2018}
R.~Neate, ``Over \$119bn wiped off facebook's market cap after growth shock,''
  Jul 2018.

\bibitem{Amazon_2016}
D.~Ingold and S.~Soper, ``Amazon doesn’t consider the race of its customers.
  should it?,'' April 2016.

\bibitem{wang2013affects}
H.-Y. Wang, C.~Liao, and L.-H. Yang, ``What affects mobile application use? the
  roles of consumption values,'' {\em International Journal of Marketing
  Studies}, vol.~5, no.~2, p.~11, 2013.

\bibitem{nissenbaum2011preemption}
H.~Nissenbaum, ``From preemption to circumvention: if technology regulates, why
  do we need regulation (and vice versa),'' {\em Berkeley Tech. LJ}, vol.~26,
  p.~1367, 2011.

\bibitem{flanagan2007game}
M.~Flanagan and H.~Nissenbaum, ``A game design methodology to incorporate
  social activist themes,'' in {\em Proceedings of the SIGCHI conference on
  Human factors in computing systems}, pp.~181--190, 2007.

\bibitem{mougouei2018operationalizing}
D.~Mougouei, H.~Perera, W.~Hussain, R.~Shams, and J.~Whittle,
  ``Operationalizing human values in software: a research roadmap,'' in {\em
  Proceedings of the 2018 26th ACM Joint Meeting on European Software
  Engineering Conference and Symposium on the Foundations of Software
  Engineering}, pp.~780--784, 2018.

\bibitem{Friedman2017}
B.~Friedman, D.~G. Hendry, and A.~Borning, ``{A survey of value sensitive
  design methods},'' {\em Foundations and Trends in Human-Computer
  Interaction}, vol.~11, no.~23, pp.~63--125, 2017.

\bibitem{Lenberg2015}
P.~Lenberg, R.~Feldt, and L.~G. Wallgren, ``{Behavioral software engineering: A
  definition and systematic literature review},'' {\em Journal of Systems and
  Software}, vol.~107, pp.~15--37, sep 2015.

\bibitem{Perera2020}
H.~Perera, W.~Hussain, J.~Whittle, A.~Nurwidyantoro, D.~Mougouei, R.~A. Shams,
  and G.~Oliver, ``{A study on the prevalence of human values in software
  engineering publications, 2015 - 2018},'' in {\em IEEE/ACM 42nd International
  Conference on Software Engineering}, pp.~409--420, 2020.

\bibitem{Salleh2019}
N.~Salleh, F.~Mendes, and E.~Mendes, ``{A Systematic Mapping Study of
  Value-Based Software Engineering},'' {\em Proceedings - 45th Euromicro
  Conference on Software Engineering and Advanced Applications, SEAA 2019},
  no.~October 2017, pp.~404--411, 2019.

\bibitem{Khurum2013}
M.~Khurum, T.~Gorschek, and M.~Wilson, ``{The software value map - An
  exhaustive collection of value aspects for the development of software
  intensive products},'' {\em Journal of software: Evolution and Process},
  vol.~25, no.~7, pp.~711--741, 2013.

\bibitem{biffl2006value}
S.~Biffl, A.~Aurum, B.~Boehm, H.~Erdogmus, and P.~Gr{\"u}nbacher, {\em
  Value-based software engineering}.
\newblock Springer Science \& Business Media, 2006.

\bibitem{webster2002analyzing}
J.~Webster and R.~T. Watson, ``Analyzing the past to prepare for the future:
  Writing a literature review,'' {\em MIS quarterly}, pp.~xiii--xxiii, 2002.

\bibitem{shams2020}
R.~A. Shams, W.~Hussain, G.~Oliver, A.~Nurwidyantoro, H.~Perera, and
  J.~Whittle, ``Society-oriented applications development: Investigating
  users’ values from bangladeshi agriculture mobile applications,'' in {\em
  Proceedings of The 42nd International Conference on Software Engineering},
  ACM, 2020.

\bibitem{schwartz1996value}
S.~Schwartz, ``Value priorities and behavior: Applying a theory of integrated
  value systems.,'' 1996.

\bibitem{gouveia2014functional}
V.~V. Gouveia, T.~L. Milfont, and V.~M. Guerra, ``Functional theory of human
  values: Testing its content and structure hypotheses,'' {\em Personality and
  Individual Differences}, vol.~60, pp.~41--47, 2014.

\bibitem{parashar2004perception}
S.~Parashar, S.~Dhar, and U.~Dhar, ``Perception of values: a study of future
  professionals,'' {\em Journal of Human Values}, vol.~10, no.~2, pp.~143--152,
  2004.

\bibitem{schwartz1987toward}
S.~H. Schwartz and W.~Bilsky, ``Toward a universal psychological structure of
  human values.,'' {\em Journal of personality and social psychology}, vol.~53,
  no.~3, p.~550, 1987.

\bibitem{rohan2000rose}
M.~J. Rohan, ``A rose by any name? the values construct,'' {\em Personality and
  social psychology review}, vol.~4, no.~3, pp.~255--277, 2000.

\bibitem{schunk2008motivation}
D.~H. Schunk, J.~R. Meece, and P.~R. Pintrich, {\em Motivation in education:
  Theory, research, and applications}.
\newblock Pearson Higher Ed, 2008.

\bibitem{tappolet2016emotions}
C.~Tappolet, {\em Emotions, values, and agency}.
\newblock Oxford University Press, 2016.

\bibitem{eccles2002motivational}
J.~S. Eccles and A.~Wigfield, ``Motivational beliefs, values, and goals,'' {\em
  Annual review of psychology}, vol.~53, no.~1, pp.~109--132, 2002.

\bibitem{fieser2016ethics}
J.~Fieser, ``Ethics. the internet encyclopedia of philosophy. issn
  2161--0002,'' 2016.

\bibitem{schwartz1992universals}
S.~H. Schwartz, ``Universals in the content and structure of values:
  Theoretical advances and empirical tests in 20 countries,'' {\em Advances in
  Experimental Social Psychology}, vol.~25, no.~1, pp.~1--65, 1992.

\bibitem{hofstede1984hofstede}
G.~Hofstede and M.~H. Bond, ``Hofstede's culture dimensions: An independent
  validation using rokeach's value survey,'' {\em Journal of cross-cultural
  psychology}, vol.~15, no.~4, pp.~417--433, 1984.

\bibitem{kahle1988using}
L.~R. Kahle and P.~Kennedy, ``Using the list of values (lov) to understand
  consumers,'' {\em Journal of Services Marketing}, 1988.

\bibitem{cheng2010developing}
A.-S. Cheng and K.~R. Fleischmann, ``Developing a meta-inventory of human
  values,'' {\em Proceedings of the American Society for Information Science
  and Technology}, vol.~47, no.~1, pp.~1--10, 2010.

\bibitem{CommonCauseHandbook}
P.~I.~R. Centre, {\em The Common Cause Handbook - A Guide to Values and Frames
  for Campaigners, Community Organisers, Civil Servants, Fundraisers,
  Educators, Social Entrepreneurs, Activists, Funders, Politicians, and
  everyone in between}.
\newblock 2011.

\bibitem{Boehm2003}
B.~Boehm and L.~G. Huang, ``{Value-Based Software Engineering: A Case Study},''
  {\em Computer}, no.~March, pp.~33--41, 2003.

\bibitem{manders2011values}
N.~Manders-Huits, ``What values in design? the challenge of incorporating moral
  values into design,'' {\em Science and engineering ethics}, vol.~17, no.~2,
  pp.~271--287, 2011.

\bibitem{mehta2019influence}
A.~Mehta, N.~P. Morris, B.~Swinnerton, and M.~Homer, ``The influence of values
  on e-learning adoption,'' {\em Computers \& Education}, vol.~141, p.~103617,
  2019.

\bibitem{zhang2020machine}
J.~M. Zhang, M.~Harman, L.~Ma, and Y.~Liu, ``Machine learning testing: Survey,
  landscapes and horizons,'' {\em IEEE Transactions on Software Engineering},
  2020.

\bibitem{mehrabi2019survey}
N.~Mehrabi, F.~Morstatter, N.~Saxena, K.~Lerman, and A.~Galstyan, ``A survey on
  bias and fairness in machine learning,'' {\em arXiv preprint
  arXiv:1908.09635}, 2019.

\bibitem{jalali2012systematic}
S.~Jalali and C.~Wohlin, ``Systematic literature studies: database searches vs.
  backward snowballing,'' in {\em Proceedings of the 2012 ACM-IEEE
  international symposium on empirical software engineering and measurement},
  pp.~29--38, IEEE, 2012.

\bibitem{kitchenham2007guidelines}
B.~Kitchenham and S.~Charters, ``Guidelines for performing systematic
  literature reviews in software engineering,'' 2007.

\bibitem{wohlin2014guidelines}
C.~Wohlin, ``Guidelines for snowballing in systematic literature studies and a
  replication in software engineering,'' in {\em Proceedings of the 18th
  international conference on evaluation and assessment in software
  engineering}, pp.~1--10, 2014.

\bibitem{wieringa2006requirements}
R.~Wieringa, N.~Maiden, N.~Mead, and C.~Rolland, ``Requirements engineering
  paper classification and evaluation criteria: a proposal and a discussion,''
  {\em Requirements engineering}, vol.~11, no.~1, pp.~102--107, 2006.

\bibitem{engstrom2011software}
E.~Engstr{\"o}m and P.~Runeson, ``Software product line testing--a systematic
  mapping study,'' {\em Information and Software Technology}, vol.~53, no.~1,
  pp.~2--13, 2011.

\bibitem{seaman1999qualitative}
C.~B. Seaman, ``Qualitative methods in empirical studies of software
  engineering,'' {\em IEEE Transactions on software engineering}, vol.~25,
  no.~4, pp.~557--572, 1999.

\bibitem{thew2018value}
S.~Thew and A.~Sutcliffe, ``Value-based requirements engineering: method and
  experience,'' {\em Requirements engineering}, vol.~23, no.~4, pp.~443--464,
  2018.

\bibitem{winter2018measuring}
E.~Winter, S.~Forshaw, and M.~A. Ferrario, ``Measuring human values in software
  engineering,'' in {\em Proceedings of the 12th ACM/IEEE International
  Symposium on Empirical Software Engineering and Measurement}, pp.~1--4, 2018.

\bibitem{zdravkovic2015capturing}
J.~Zdravkovic, E.-O. Svee, and C.~Giannoulis, ``Capturing consumer preferences
  as requirements for software product lines,'' {\em Requirements engineering},
  vol.~20, no.~1, pp.~71--90, 2015.

\bibitem{pereira2015value}
R.~Pereira and M.~C.~C. Baranauskas, ``A value-oriented and culturally informed
  approach to the design of interactive systems,'' {\em International Journal
  of Human-Computer Studies}, vol.~80, pp.~66--82, 2015.

\bibitem{barn2015role}
B.~Barn, R.~Barn, and F.~Raimondi, ``On the role of value sensitive concerns in
  software engineering practice,'' in {\em 2015 IEEE/ACM 37th IEEE
  International Conference on Software Engineering}, vol.~2, pp.~497--500,
  IEEE, 2015.

\bibitem{aldewereld2015design}
H.~Aldewereld, V.~Dignum, and Y.-h. Tan, ``Design for values in software
  development,'' {\em Handbook of Ethics, Values, and Technological Design:
  Sources, theory, values and application domains, Berlin: Springer}, 2015.

\bibitem{pereira2014value}
R.~Pereira and M.~C.~C. Baranauskas, ``Value pie: a culturally informed
  conceptual scheme for understanding values in design,'' in {\em International
  Conference on Human-Computer Interaction}, pp.~122--133, Springer, 2014.

\bibitem{anthonysamy2012method}
P.~Anthonysamy, P.~Greenwood, and A.~Rashid, ``A method for analysing
  traceability between privacy policies and privacy controls of online social
  networks,'' in {\em Annual Privacy Forum}, pp.~187--202, Springer, 2012.

\bibitem{sjokvist2019eliciting}
N.~M. Sj{\o}kvist and M.~Kj{\o}rstad, ``Eliciting human values by applying
  design thinking techniques in systems engineering,'' in {\em INCOSE
  International Symposium}, vol.~29, pp.~478--499, Wiley Online Library, 2019.

\bibitem{yoo2013value}
D.~Yoo, A.~Huldtgren, J.~P. Woelfer, D.~G. Hendry, and B.~Friedman, ``A value
  sensitive action-reflection model: evolving a co-design space with
  stakeholder and designer prompts,'' in {\em Proceedings of the SIGCHI
  conference on human factors in computing systems}, pp.~419--428, 2013.

\bibitem{iversen2012values}
O.~S. Iversen, K.~Halskov, and T.~W. Leong, ``Values-led participatory
  design,'' {\em CoDesign}, vol.~8, no.~2-3, pp.~87--103, 2012.

\bibitem{colomo2011using}
R.~Colomo-Palacios, C.~Casado-Lumbreras, P.~Soto-Acosta, and
  {\'A}.~Garc{\'\i}a-Crespo, ``Using the affect grid to measure emotions in
  software requirements engineering,'' 2011.

\bibitem{detweiler2010principles}
C.~Detweiler, K.~Hindriks, and C.~Jonker, ``Principles for value-sensitive
  agent-oriented software engineering,'' in {\em International Workshop on
  Agent-Oriented Software Engineering}, pp.~1--16, Springer, 2010.

\bibitem{proynova2011investigating}
R.~Proynova, B.~Paech, S.~H. Koch, A.~Wicht, and T.~Wetter, ``Investigating the
  influence of personal values on requirements for health care information
  systems,'' in {\em Proceedings of the 3rd Workshop on Software Engineering in
  Health Care}, pp.~48--55, 2011.

\bibitem{li2009bridge}
Q.~Li, M.~Li, Y.~Yang, Q.~Wang, T.~Tan, B.~Boehm, and C.~Hu, ``Bridge the gap
  between software test process and business value: a case study,'' in {\em
  International Conference on Software Process}, pp.~212--223, Springer, 2009.

\bibitem{perera2020continual}
H.~Perera, G.~Mussbacher, W.~Hussain, R.~A. Shams, A.~Nurwidyantoro, and
  J.~Whittle, ``Continual human value analysis in software development: A goal
  model based approach,'' in {\em 2020 IEEE 28th International Requirements
  Engineering Conference (RE)}, pp.~192--203, IEEE, 2020.

\bibitem{nathan2008envisioning}
L.~P. Nathan, B.~Friedman, P.~Klasnja, S.~K. Kane, and J.~K. Miller,
  ``Envisioning systemic effects on persons and society throughout interactive
  system design,'' in {\em Proceedings of the 7th ACM conference on Designing
  interactive systems}, pp.~1--10, 2008.

\bibitem{ramos2005requirements}
I.~Ramos, D.~M. Berry, and J.~{\'A}. Carvalho, ``Requirements engineering for
  organizational transformation,'' {\em Information and Software Technology},
  vol.~47, no.~7, pp.~479--495, 2005.

\bibitem{miller2005agile}
K.~W. Miller and D.~K. Larson, ``Agile software development: human values and
  culture,'' {\em IEEE Technology and Society Magazine}, vol.~24, no.~4,
  pp.~36--42, 2005.

\bibitem{flanagan2005values}
M.~Flanagan, D.~C. Howe, and H.~Nissenbaum, ``Values at play: Design tradeoffs
  in socially-oriented game design,'' in {\em Proceedings of the SIGCHI
  conference on human factors in computing systems}, pp.~751--760, 2005.

\bibitem{cockton2005development}
G.~Cockton, ``A development framework for value-centred design,'' in {\em
  CHI'05 extended abstracts on Human factors in computing systems},
  pp.~1292--1295, 2005.

\bibitem{mussbacher2020there}
G.~Mussbacher, W.~Hussain, and J.~Whittle, ``Is there a need to address human
  values in domain modelling?,'' in {\em 2020 IEEE Tenth International
  Model-Driven Requirements Engineering (MoDRE)}, pp.~73--77, IEEE, 2020.

\bibitem{surian2011recommending}
D.~Surian, N.~Liu, D.~Lo, H.~Tong, E.-P. Lim, and C.~Faloutsos, ``Recommending
  people in developers' collaboration network,'' in {\em 2011 18th Working
  Conference on Reverse Engineering}, pp.~379--388, IEEE, 2011.

\bibitem{lee2014customer}
Y.~K. Lee, H.~P. In, and R.~Kazman, ``Customer requirements validation method
  based on mental models,'' in {\em 2014 21st Asia-Pacific Software Engineering
  Conference}, vol.~1, pp.~199--206, IEEE, 2014.

\bibitem{schuler2013rule}
A.~Schuler and B.~Franz, ``Rule-based generation of mobile user interfaces,''
  in {\em 2013 10th International Conference on Information Technology: New
  Generations}, pp.~267--272, IEEE, 2013.

\bibitem{amreen2019developer}
S.~Amreen, A.~Karnauch, and A.~Mockus, ``Developer reputation estimator
  (dre),'' in {\em 2019 34th IEEE/ACM International Conference on Automated
  Software Engineering (ASE)}, pp.~1082--1085, IEEE, 2019.

\bibitem{rathnayake2019framework}
N.~Rathnayake, D.~Meedeniya, I.~Perera, and A.~Welivita, ``A framework for
  adaptive user interface generation based on user behavioural patterns,'' in
  {\em 2019 Moratuwa Engineering Research Conference (MERCon)}, pp.~698--703,
  IEEE, 2019.

\bibitem{romero2019adapting}
V.~Romero-Chac{\'o}n, H.~Muir-Camacho, J.~Rodr{\'\i}guez-Gonz{\'a}lez,
  A.~G{\'o}mez-Blanco, and M.~Chac{\'o}n-Rivas, ``Adapting scrum methodology to
  develop accessible web sites,'' in {\em 2019 International Conference on
  Inclusive Technologies and Education (CONTIE)}, pp.~112--1124, IEEE.

\bibitem{barivsic2017requirements}
A.~Bari{\v{s}}i{\'c}, D.~Blouin, V.~Amaral, and M.~Goul{\~a}o, ``A requirements
  engineering approach for usability-driven dsl development,'' in {\em
  Proceedings of the 10th ACM SIGPLAN International Conference on Software
  Language Engineering}, pp.~115--128, 2017.

\bibitem{albarghouthi2019fairness}
A.~Albarghouthi and S.~Vinitsky, ``Fairness-aware programming,'' in {\em
  Proceedings of the Conference on Fairness, Accountability, and Transparency},
  pp.~211--219, 2019.

\bibitem{ying2016earec}
H.~Ying, L.~Chen, T.~Liang, and J.~Wu, ``Earec: leveraging expertise and
  authority for pull-request reviewer recommendation in github,'' in {\em 2016
  IEEE/ACM 3rd International Workshop on CrowdSourcing in Software Engineering
  (CSI-SE)}, pp.~29--35, IEEE, 2016.

\bibitem{pellegrini2019prioritize}
F.~Pellegrini, M.~Anjos, F.~Florentin, B.~Ribeiro, W.~Correia, and J.~Quintino,
  ``How to prioritize accessibility in agile projects,'' in {\em International
  Conference on Applied Human Factors and Ergonomics}, pp.~271--280, Springer,
  2019.

\bibitem{de2020moderating}
C.~R. de~Souza, L.~S. Machado, and R.~R.~M. Melo, ``On moderating software
  crowdsourcing challenges,'' {\em Proceedings of the ACM on Human-Computer
  Interaction}, vol.~4, no.~GROUP, pp.~1--22, 2020.

\bibitem{curumsing2019emotion}
M.~K. Curumsing, N.~Fernando, M.~Abdelrazek, R.~Vasa, K.~Mouzakis, and
  J.~Grundy, ``Emotion-oriented requirements engineering: A case study in
  developing a smart home system for the elderly,'' {\em Journal of Systems and
  Software}, vol.~147, pp.~215--229, 2019.

\bibitem{harbers2015embedding}
M.~Harbers, C.~Detweiler, and M.~A. Neerincx, ``Embedding stakeholder values in
  the requirements engineering process,'' in {\em International Working
  Conference on Requirements Engineering: Foundation for Software Quality},
  pp.~318--332, Springer, 2015.

\bibitem{pedell2015don}
S.~Pedell, A.~A. Lopez-Lorca, T.~Miller, and L.~Sterling, ``Don't leave me
  untouched: considering emotions in personal alarm use and development,'' in
  {\em Healthcare Informatics and Analytics: Emerging Issues and Trends},
  pp.~96--127, IGI Global, 2015.

\bibitem{kheirandish2019huvalue}
S.~Kheirandish, M.~Funk, S.~Wensveen, M.~Verkerk, and M.~Rauterberg, ``Huvalue:
  a tool to support design students in considering human values in their
  design,'' {\em International Journal of Technology and Design Education},
  pp.~1--27, 2019.

\bibitem{giorgini2006requirements}
P.~Giorgini, F.~Massacci, J.~Mylopoulos, and N.~Zannone, ``Requirements
  engineering for trust management: model, methodology, and reasoning,'' {\em
  International Journal of Information Security}, vol.~5, no.~4, pp.~257--274,
  2006.

\bibitem{uddin2008umltrust}
M.~G. Uddin and M.~Zulkernine, ``Umltrust: towards developing trust-aware
  software,'' in {\em Proceedings of the 2008 ACM symposium on Applied
  computing}, pp.~831--836, 2008.

\bibitem{doerr2007built}
J.~Doerr, S.~Hartkopf, D.~Kerkow, D.~Landmann, and P.~Amthor, ``Built-in user
  satisfaction-feature appraisal and prioritization with amuse,'' in {\em 15th
  IEEE International Requirements Engineering Conference (RE 2007)},
  pp.~101--110, IEEE, 2007.

\bibitem{mchugh2011agile}
O.~McHugh, K.~Conboy, and M.~Lang, ``Agile practices: The impact on trust in
  software project teams,'' {\em Ieee Software}, vol.~29, no.~3, pp.~71--76,
  2011.

\bibitem{aydemir2018roadmap}
F.~B. Aydemir and F.~Dalpiaz, ``A roadmap for ethics-aware software
  engineering,'' in {\em 2018 IEEE/ACM International Workshop on Software
  Fairness (FairWare)}, pp.~15--21, IEEE, 2018.

\bibitem{lopez2019talking}
T.~Lopez, H.~Sharp, T.~Tun, A.~Bandara, M.~Levine, and B.~Nuseibeh, ``Talking
  about security with professional developers,'' in {\em 2019 IEEE/ACM Joint
  7th International Workshop on Conducting Empirical Studies in Industry (CESI)
  and 6th International Workshop on Software Engineering Research and
  Industrial Practice (SER\&IP)}, pp.~34--40, IEEE, 2019.

\bibitem{sutcliffe2006pc}
A.~Sutcliffe, S.~Fickas, and M.~M. Sohlberg, ``Pc-re: a method for personal and
  contextual requirements engineering with some experience,'' {\em Requirements
  Engineering}, vol.~11, no.~3, pp.~157--173, 2006.

\bibitem{liaskos2011representing}
S.~Liaskos, S.~A. McIlraith, S.~Sohrabi, and J.~Mylopoulos, ``Representing and
  reasoning about preferences in requirements engineering,'' {\em Requirements
  Engineering}, vol.~16, no.~3, p.~227, 2011.

\bibitem{tramer2017fairtest}
F.~Tramer, V.~Atlidakis, R.~Geambasu, D.~Hsu, J.-P. Hubaux, M.~Humbert,
  A.~Juels, and H.~Lin, ``Fairtest: Discovering unwarranted associations in
  data-driven applications,'' in {\em 2017 IEEE European Symposium on Security
  and Privacy (EuroS\&P)}, pp.~401--416, IEEE, 2017.

\bibitem{sindre2005eliciting}
G.~Sindre and A.~L. Opdahl, ``Eliciting security requirements with misuse
  cases,'' {\em Requirements engineering}, vol.~10, no.~1, pp.~34--44, 2005.

\bibitem{mougouei2020engineering}
D.~Mougouei, ``Engineering human values in software through value
  programming,'' in {\em Proceedings of the IEEE/ACM 42nd International
  Conference on Software Engineering Workshops}, pp.~133--136, 2020.

\bibitem{aksnes2003characteristics}
D.~W. Aksnes, ``Characteristics of highly cited papers,'' {\em Research
  evaluation}, vol.~12, no.~3, pp.~159--170, 2003.

\bibitem{Sommerville2016}
I.~Sommerville, {\em Software Engineering}.
\newblock Pearson education, 2016.

\bibitem{friedman2012envisioning}
B.~Friedman and D.~Hendry, ``The envisioning cards: a toolkit for catalyzing
  humanistic and technical imaginations,'' in {\em Proceedings of the SIGCHI
  conference on human factors in computing systems}, pp.~1145--1148, 2012.

\bibitem{dindler2007fictional}
C.~Dindler and O.~S. Iversen, ``Fictional inquiry—design collaboration in a
  shared narrative space,'' {\em CoDesign}, vol.~3, no.~4, pp.~213--234, 2007.

\bibitem{9261980}
W.~{Hussain}, H.~{Perera}, J.~{Whittle}, A.~{Nurwidyantoro}, R.~{Hoda}, R.~A.
  {Shams}, and G.~{Oliver}, ``Human values in software engineering: Contrasting
  case studies of practice,'' {\em IEEE Transactions on Software Engineering},
  pp.~1--1, 2020.

\bibitem{213081}
A.~{Dardenne}, S.~{Fickas}, and A.~{van Lamsweerde}, ``Goal-directed concept
  acquisition in requirements elicitation,'' in {\em Proceedings of the Sixth
  International Workshop on Software Specification and Design}, pp.~14--21,
  1991.

\bibitem{martin1983human}
T.~Martin, ``Human software requirements engineering for computer-controlled
  manufacturing systems,'' in {\em Analysis, Design and Evaluation of
  Man--Machine Systems}, pp.~151--156, Elsevier, 1983.

\bibitem{vigo2013benchmarking}
M.~Vigo, J.~Brown, and V.~Conway, ``Benchmarking web accessibility evaluation
  tools: measuring the harm of sole reliance on automated tests,'' in {\em
  Proceedings of the 10th International Cross-Disciplinary Conference on Web
  Accessibility}, pp.~1--10, 2013.

\bibitem{story2021awareness}
P.~Story, D.~Smullen, Y.~Yao, A.~Acquisti, L.~F. Cranor, N.~Sadeh, and
  F.~Schaub, ``Awareness, adoption, and misconceptions of web privacy tools,''
  {\em Proceedings on Privacy Enhancing Technologies}, vol.~3, pp.~308--333,
  2021.

\bibitem{rogers2010diffusion}
E.~M. Rogers, {\em Diffusion of innovations}.
\newblock Simon and Schuster, 2010.

\bibitem{verbeek2008morality}
P.-P. Verbeek, ``Morality in design: Design ethics and the morality of
  technological artifacts,'' in {\em Philosophy and design}, pp.~91--103,
  Springer, 2008.

\bibitem{crawford2016artificial}
K.~Crawford, ``Artificial intelligence’s white guy problem,'' {\em The New
  York Times}, vol.~25, no.~06, 2016.

\bibitem{flanagan2014values}
M.~Flanagan and H.~Nissenbaum, {\em Values at play in digital games}.
\newblock MIT Press, 2014.

\bibitem{winner1980artifacts}
L.~Winner, ``Do artifacts have politics?,'' {\em Daedalus}, pp.~121--136, 1980.

\bibitem{winter2019advancing}
E.~Winter, S.~Forshaw, L.~Hunt, and M.~A. Ferrario, ``Advancing the study of
  human values in software engineering,'' in {\em 2019 IEEE/ACM 12th
  International Workshop on Cooperative and Human Aspects of Software
  Engineering (CHASE)}, pp.~19--26, IEEE, 2019.

\bibitem{hailpern2002software}
B.~Hailpern and P.~Santhanam, ``Software debugging, testing, and
  verification,'' {\em IBM Systems Journal}, vol.~41, no.~1, pp.~4--12, 2002.

\bibitem{farooq2011software}
S.~U. Farooq, S.~Quadri, and N.~Ahmad, ``Software measurements and metrics:
  Role in effective software testing,'' {\em International Journal of
  Engineering Science and Technology}, vol.~3, no.~1, pp.~671--680, 2011.

\bibitem{garousi2016and}
V.~Garousi and M.~V. M{\"a}ntyl{\"a}, ``When and what to automate in software
  testing? a multi-vocal literature review,'' {\em Information and Software
  Technology}, vol.~76, pp.~92--117, 2016.

\end{thebibliography}
\begin{IEEEbiography}[{\includegraphics[width=1in,height=1.25in,clip,keepaspectratio]{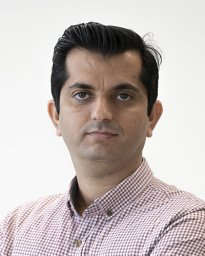}}]{Mojtaba Shahin} is a Software Engineering Lecturer at RMIT University, Australia. Previously, he was a Research Fellow at Monash University. His research interests reside in Empirical Software Engineering, Human and Social Aspects of Software Engineering, and Secure Software Engineering. He completed his PhD study at the University of Adelaide, Australia.
\end{IEEEbiography}

\begin{IEEEbiography}[{\includegraphics[width=1in,height=1.25in,clip,keepaspectratio]{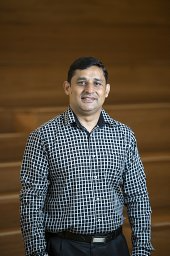}}]{Waqar Hussain} is a Senior Research Scientist at CSIRO's Data61. His research interests include ethical artificial intelligence, values-based systems development, human-centric software engineering, and empirical software engineering.
\end{IEEEbiography}

\begin{IEEEbiography}[{\includegraphics[width=1in,height=1.25in,clip,keepaspectratio]{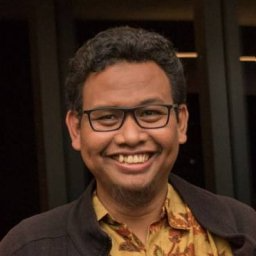}}]{Arif Nurwidyantoro} is a PhD student at Monash University. He received his BSc (Hons.) degree from Institut Pertanian Bogor, Indonesia, and the MSc degree from Universitas Gadjah Mada, Indonesia. His research interests include data analytics and software engineering.
\end{IEEEbiography}

\begin{IEEEbiography}[{\includegraphics[width=1in,height=1.25in,clip,keepaspectratio]{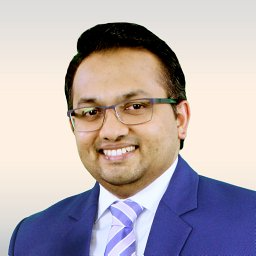}}]{Harsha Perera} is a PhD student at Monash University. He received his BSc degree from the University of Colombo, Sri Lanka, in 2015. His research interests include the intersection of Human Values and Software Engineering.
\end{IEEEbiography}

\begin{IEEEbiography}[{\includegraphics[width=1in,height=1.25in,clip,keepaspectratio]{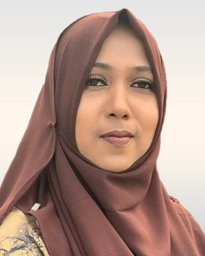}}]{Rifat Shams} is a PhD student at Monash University. She received her BSc (Hons) and master’s degrees from the University of Rajshahi, Bangladesh. Her research interests include human values-centric software development. To be very specific, her PhD is on operationalizing human values in mobile applications.
\end{IEEEbiography}

\begin{IEEEbiography}[{\includegraphics[width=1in,height=1.25in,clip,keepaspectratio]{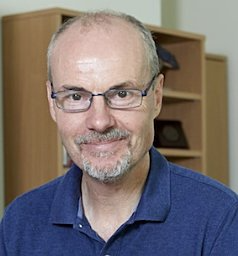}}]{John Grundy} is an Australian Laureate Fellow and Professor of Software Engineering at Monash University. He leads the HumaniSE research lab. Currently, he researches new approaches to engineering software systems that fully take into account the "human" aspects of end-users and team members.
\end{IEEEbiography}

\begin{IEEEbiography}[{\includegraphics[width=1in,height=1.25in,clip,keepaspectratio]{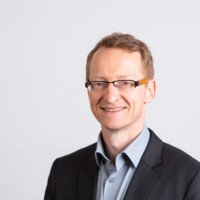}}]{Jon Whittle} is director of CSIRO’s Data61, the digital technologies and data science arm of Australia’s national science agency. He is also an adjunct (full) professor with the Faculty of Information Technology, Monash University, Melbourne. His research interests include the intersection of software engineering and human-computer interaction. He is best known for his work in model-driven development, aspect-oriented modelling, digital technologies for social good, and values in software.
\end{IEEEbiography}

\EOD
\end{document}